\documentclass[%
 reprint,
superscriptaddress,
 amsmath,amssymb,
 aps,
]{revtex4-2}

\usepackage[utf8]{inputenc}
\usepackage[T1]{fontenc}
\usepackage{caption}
\usepackage{subcaption}
\usepackage{xcolor}

\usepackage{graphicx}
\usepackage{dcolumn}
\usepackage{bm}
\usepackage{hyperref}

\usepackage{booktabs}
\usepackage{multirow}

\usepackage{algorithm}
\usepackage{algpseudocode}
\newif\ifsubmit

\submitfalse
\ifsubmit

    \newcommand{\moh}[1]{}
    \newcommand{\dima}[1]{}
    \newcommand{\dmitrii}[1]{}
    \newcommand{\johnpaul}[1]{}
    \newcommand{\todo}[1]{}
    \newcommand{\tocite}[1]{}
    
    \newcommand{\deletion}[1]{}

\else
    
    \definecolor{comments}{rgb}{0.3, 0.7, 1.0}
    \newcommand{\moh}[1]{[{\color{comments}Fuxi: #1}]}
    \newcommand{\dima}[1]{[{\color{comments}Suhas: #1}]}
    \newcommand{\dmitrii}[1]{[{\color{comments}Thomas: #1}]}
    \newcommand{\johnpaul}[1]{[{\color{comments}John Paul: #1}]}
    \newcommand{\todo}[1]{[{\color{red}TODO: #1}]}
    \newcommand{\tocite}[1]{[{\color{red}CITE: #1}]}
    
    \newcommand{\deletion}[1]{}

\fi

\begin{document}

\preprint{APS/123-QED}
\title{Energy landscapes of combinatorial optimization in Ising machines}


\author{Dmitrii Dobrynin}
\email{d.dobrynin@fz-juelich.de}
\affiliation{
    Peter Grünberg Institut (PGI-14), Forschungszentrum Jülich GmbH, Jülich, Germany
}
\affiliation{
    RWTH Aachen University, Aachen, Germany
}

\author{Adrien Renaudineau}
\affiliation{
    Peter Grünberg Institut (PGI-14), Forschungszentrum Jülich GmbH, Jülich, Germany
}

\author{Mohammad Hizzani}
\affiliation{
    Peter Grünberg Institut (PGI-14), Forschungszentrum Jülich GmbH, Jülich, Germany
}
\affiliation{
    RWTH Aachen University, Aachen, Germany
}

\author{\\Dmitri Strukov}
\affiliation{
    University of California Santa Barbara, Santa Barbara, CA, USA
}

\author{Masoud Mohseni}
\affiliation{
    LSIP, Hewlett Packard Labs, Milpitas, CA, USA
}

\author{John Paul Strachan}
\email{j.strachan@fz-juelich.de}
\affiliation{
    Peter Grünberg Institut (PGI-14), Forschungszentrum Jülich GmbH, Jülich, Germany
}
\affiliation{
    RWTH Aachen University, Aachen, Germany
}

\date{\today}

\begin{abstract}
Physics-based Ising machines (IM) have been developed as dedicated processors 
for solving hard combinatorial optimization problems with higher speed and better energy efficiency. 
Generally, such systems employ local search heuristics 
to traverse energy landscapes in searching for optimal solutions. 
Here, we quantify and address some of the major challenges 
met by IMs by extending energy-landscape geometry visualization tools known as \textit{disconnectivity graphs}. Using efficient sampling methods, we visually capture landscapes
of problems having diverse structure and hardness  manifesting as energetic and entropic barriers for IMs.  
We investigate energy barriers, local minima, and configuration space clustering effects 
caused by locality reduction methods when embedding combinatorial problems to the Ising hardware. 
To this end, we sample disconnectivity graphs of
PUBO energy landscapes and their different QUBO mappings 
accounting for both local minima and saddle regions.
We demonstrate that QUBO energy landscape properties lead to the 
subpar performance for quadratic IMs and suggest directions for their improvement.

\end{abstract}

\maketitle


\section{Introduction}
Recent years have seen an increasing interest in using classical and quantum Ising machines (IM)
for solving combinatorial optimization problems relevant for fundamental research and industrial 
applications \cite{mohseni2022}. Most of these devices rely on algorithms and physical principles 
implementing heuristic local search routines, e.g. discrete Monte-Carlo (MC) sampling (simulated annealing, 
parallel tempering) or noisy/chaotic continuous dynamics. Examples of the former are memristive 
crossbar arrays employed to efficiently perform vector-matrix multiplication \cite{mahmoodi2019, cai2020}, 
or digital ASIC annealers \cite{aramon2019}. 
The latter versions of IMs include coherent Ising machines, oscillator networks, 
quantum annealers, and others \cite{peng2008, borders2019, mallick2020, goto2021}. The main attraction for the use of IM is 
the intrinsic compatibility of the algorithm operations with their physical implementations, which
offers reducing 
time-to-solution and/or energy-to-solution metrics polynomially, or by a significant pre-factor \cite{hamerly2019, patel2020, leleu2021scaling, kowalsky2022}.

In this context, there are several outstanding challenges faced by IMs on both algorithmic and 
hardware levels, resulting in strong compromises being adopted in their practical deployment. One
and possibly the most important difficulty concerns their application to practically interesting (large) 
problem sizes. The support of only second-order couplings of ``spins'', together with connectivity topology 
constraints (e.g. the chimera graph~\cite{konz2021}) results in the introduction of multiple auxiliary
variables in order to either avoid higher-order terms, or reach necessary levels of sparsity. 
The added new variables can scale super-linearly in the number of original variables, 
not only further challenging the scaling to large problems, but also increasing 
the search space and modifying the optimization energy landscape. As a result, IMs can be limited to smaller-scale 
problems, and even these can become harder than their native formulation \cite{perdomo2019, valiante2021, hizzani2023, bybee2023} due to 
the worsened landscape geometry.

A second challenge lies in the algorithmic limitations of IMs. In particular, their reliance on local 
search heuristics fundamentally puts a bound on the problem classes they are
capable of solving \cite{gamarnik2021}. Being inherently local, 
IMs are prone to suffer from
energy barriers rejecting MC moves, and from entropic barriers or degeneracies
hampering both sensible exploration and exploitation \cite{bernaschi2021}.  However, Nonlocal Monte Carlo algorithms have been recently proposed that could significantly accelerate exploration by unmasking certain underlying structures in the configuration space \cite{mohseni2021}. 
Clear understanding of geometrical or energy landscape features of benchmark problems and the corresponding constraints of the Ising 
hardware is essential to facilitate future advances in the field.

A major challenge for designing discrete optimization/sampling solvers is the lack of understanding or representation of the high-dimensional 
configuration space. 
Only a few methods have been developed over the years to visualize high-dimensional
cost/energy functions of such problems. One example is 
disconnectivity graphs (DG, also called barrier trees) \cite{becker1997, wales1998, doye1999}, 
which aim to simplify the exponentially large 
configuration space by capturing local minima and their connectivity through energy barriers. 
It is possible to use DGs to gain quantitative insights into phenomena in a variety of applications 
ranging from metastable states of protein folding \cite{chakraborty2014}
to thermodynamic effects in Lennard-Jones systems \cite{calvo2007}, biomolecules \cite{wales2005}, and 
spin glasses \cite{garstecki1999}. However, due to exponential complexity of DG construction and high degeneracy 
of the solution space, attaining energy landscape visualization
is a computational feat on its own~\cite{biswas2020}.

The contribution of this paper is as follows. Firstly, in Sec.~\ref{sec:methods} we describe an extension for the
efficient sampling algorithm of \cite{zhou2011} to support DGs of energy landscapes featuring 
strong degeneracy of the configuration space (millions of states), capturing not only local minima but also  saddle regions. 
We further modify this approach providing means 
to construct DGs for quadratic optimization problems resulting from locality reduction due 
to IM hardware mapping. We achieve this by meaningfully reducing the search space over auxiliary 
variables and defining ``effective'' barriers. 
Secondly, in Sec.~\ref{sec:easy_and_hard_dg} and Sec.~\ref{sec:rand_industrial_dg} 
using 3-SAT as a representative higher-order
problem class, we plot DGs for problems of sizes 
inaccessible to the methods reported previously. With our methods
we compare easy to hard instances, and random to industrial (structured) instances.
Finally, in Sec.~\ref{sec:qubo_3sat} we demonstrate suboptimal energy landscape features of 
hardware embedding quadratization methods for 3-SAT 
from the perspective of clustering and entropy of energy minima, 
which are some of the culprits of algorithmic hardness \cite{coja2017, bellitti2021}.

\section{Background \label{sec:background}}
The conventional (2-local) Ising Hamiltonian, which IMs natively solve is:
\begin{equation}
    H_{\mathrm{Ising}} = \sum_{i < j}^{N}J_{ij}s_i s_j + \sum_{i}^N h_is_i\,,
    \label{eq:Ising_H}
\end{equation}
where $s_i \in \{-1, 1\}$, $J_{ij}$ are spin interaction strengths, and $h_i$ denote local magnetic fields.
Finding the ground state of Eq.~\ref{eq:Ising_H} is an NP-Complete problem \cite{barahona1982}, and therefore 
approximately solving this Hamiltonian efficiently is of profound interest. Alternatively, the Ising 
Hamiltonian can be formulated as a quadratic pseudo-boolean function:
\begin{equation}
    H_{\mathrm{QUBO}} = \sum_{i < j}^{N}Q_{ij}x_i x_j + \sum_{i}^N b_ix_i + C\,,
    \label{eq:QUBO_H}
\end{equation}
with binary variables $x_i \in \{0, 1\}$. Deciding the ground state of 
this function among $2^N$ possible configurations 
is commonly called 
Quadratic Unconstrained Binary Optimization (QUBO) problem.
In this work, we will use Ising and QUBO terms interchangeably due to their equivalence.

The generalization of QUBO to support higher order interactions of variables is usually referred to as PUBO (``P'' for polynomial):
\begin{equation}
    f(\mathbf{x}) = f(x_1, x_2, \dots, x_N)= \sum_{\{i\}_k\subseteq V}a_{\{i\}}\prod_{\{i\}_k}x_i + C\,,
    \label{eq:PUBO}
\end{equation}
which is correspondingly equivalent to the k-local Ising (historically called the p-order Ising spin glass \cite{gardner1985}). 
Here $\{i\}_k\subseteq V$ denote all possible subsets of the set of variables 
with the order of interaction not larger than the highest $k\ge 1$.

The present work devotes particular attention to the k-SAT problem (see below Eq.~\ref{eq:k_SAT}),
one of the oldest and well-studied NP-Complete problems \cite{selman1996, mezard2005a, mezard2009}. 
The motivation behind this choice lies in 
the fact that apart from being practically important for a variety of applications 
\cite{silva2008}, k-SAT highlights the hardware and algorithmic challenges of IMs \cite{bhattacharya2024}. 
As will be discussed in this work, it features strong degeneracy of the solution space, an abundance 
of energy barriers, clustering of solutions, and can only be natively supported by the PUBO formulation, 
making it a formidable problem class for local search based quadratic IMs. 

A general statement of the k-SAT decision problem is simple: 
is there a binary variable assignment $\mathbf{x} \in \mathbb{B}^N$ of the following 
conjunctive normal form (CNF):
\begin{equation}
    (l_{i_{1,1}} \vee l_{i_{1,2}} \dots \vee l_{i_{1,k}})\wedge \dots\wedge(l_{i_{m,1}} \vee l_{i_{m,2}} \dots \vee l_{i_{m,k}}) \,,
    \label{eq:k_SAT}
\end{equation}
where $i \in \{1, N\}$, $m \in \{1, M\}$, $l = x$ or $l = \bar{x}$, so that all $M$ clauses are satisfied?
With $k\ge 3$ it is NP-Complete like Ising/QUBO and thus worst case exponentially hard \cite{garey1990}.
The k-local PUBO cost function Eq.~\ref{eq:PUBO} is easily obtained from Eq.~\ref{eq:k_SAT} 
as shown below in methods by Eq.~\ref{eq:pubo_from_cnf}.

Many  optimization landscape features have been established for hard constraint satisfaction problems 
\cite{mezard2005a, montanari2008}, of which k-SAT is a conventional example. 
By increasing the number of constraints from the small number, where the problem is easily 
satisfiable, to larger values up to a point of unsatisfiability, optimization landscapes undergo
phase transitions where the dominating ``simple'' configuration region of connected  
solutions gets shattered into exponentially many \textit{clusters} of solutions. Each cluster consists of 
several configurations which can be easily accessed from each other by local dynamics \cite{zdeborova2007}. 
Furthermore, some of the variables in such cluster configurations could also be ``frozen'' \cite{ardelius2008, li2009}, 
i.e. remain unchanged regardless of the state of others. In other words, not only it can be difficult 
to traverse the landscape in the search of isolated clusters, but also to transition between such clusters, 
it is imperative to modify an extensive fraction of variables simultaneously; 
thus, \textit{non-local} moves can be essential \cite{mohseni2021}.  
Recently, there has been renewed interest to quantify geometrical aspects of the 
algorithmic hardness near a computational phase transition by introducing the notion of Overlap Gap Properties (OGP)
\cite{bresler2021, gamarnik2021, gamarnik2022}. 
In order to illustrate energy landscape geometry features as a cause of hardness of combinatorial optimization in IMs, 
in this work we focus on illustrating how the landscapes are perceived by \textit{local} search.

\begin{figure}[ht]
    \centering
    \includegraphics[width=0.9\linewidth]{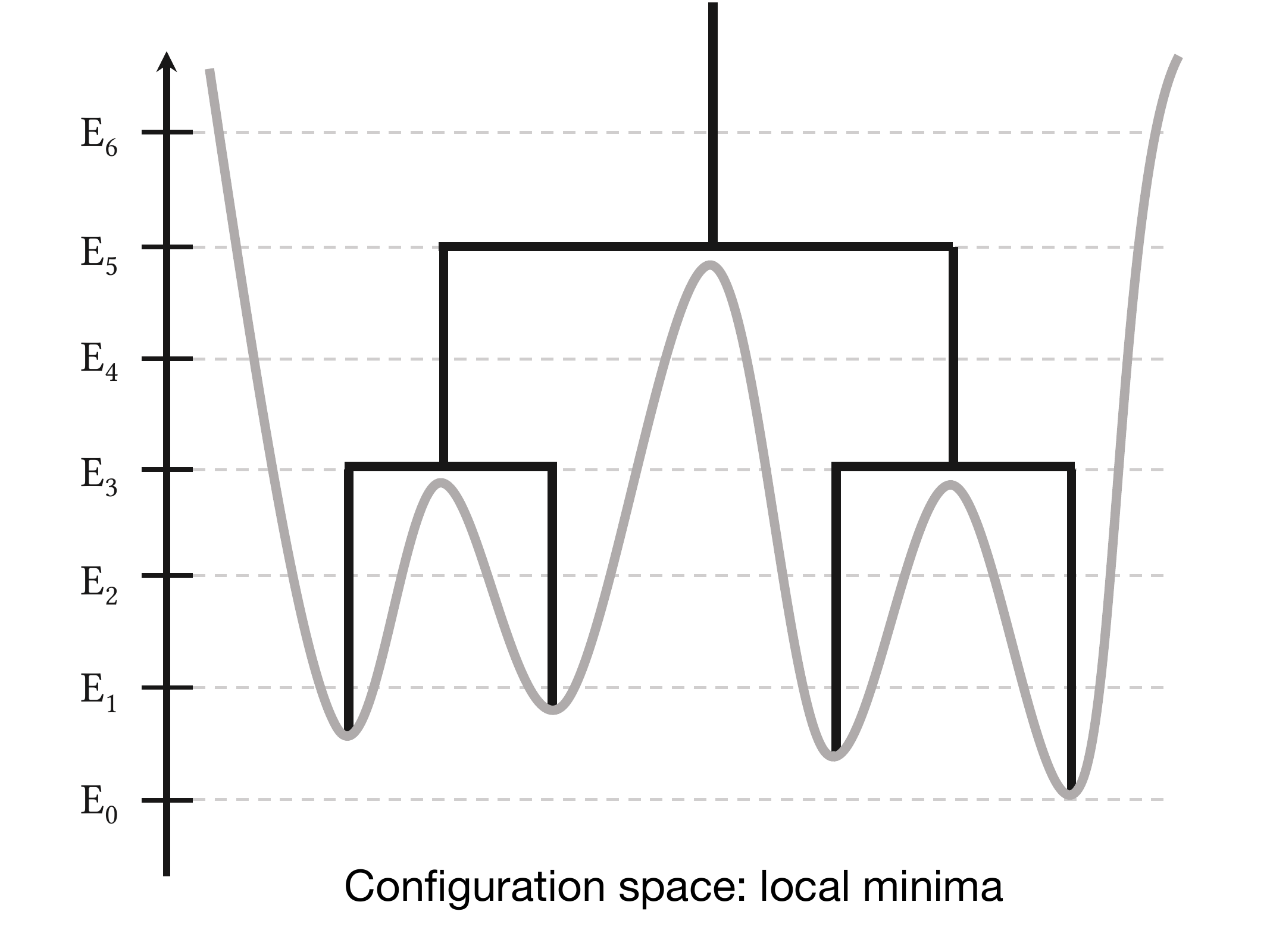}
    \caption{A simplified view of a disconnectivity graph. Every local minimum 
        corresponds to a leaf; the height of energy barriers is reflected by the energy of 
        branch connections. Horizontal arrangement of minima \textit{does not} represent 
        distance, i.e. by default has no explicit meaning.}
    \label{fig:DG}
\end{figure}

Early efforts to visualize energy/fitness landscapes arose in the context of theoretical chemistry 
and biology \cite{becker1997, wales1998, doye1999}. Authors of these works introduced the concept of 
disconnectivity graphs (DG) implementing a map of exponentially large potential energy configuration spaces 
to a two-dimensional tree. 
Fig.~\ref{fig:DG} sketches the idea behind such mapping: 
every leaf corresponds to a local minimum, while the branches represent the magnitude of 
energy barriers and connectivity (lowest barrier separation) of local minima with respect to each other.

In principle, arbitrary energy landscapes can be defined by a triplet \cite{reidys2002}:
$X$ being a set of configurations, neighbourhood $\mathcal{N}(x)$ of every state $x$ in $X$, 
and energy/fitness function $f(X) \in \mathbb{R}$. We say that a solver explores the energy landscape if a 
local search move from any given configuration $x$ chooses a state in $\mathcal{N}(x)$.
For instance, one may choose a random neighbour (random walk) or the one with the largest energy decrease 
(steepest descent).

\begin{figure}[ht]
     \centering
     \begin{subfigure}[b]{0.42\linewidth}
         \centering
         \includegraphics[width=\linewidth]{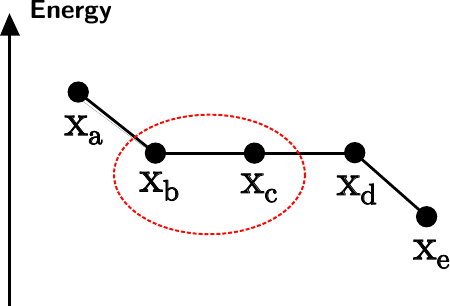}
         \caption{}
         \label{fig:shoulder}
     \end{subfigure}
     \hspace{0.02\linewidth}
     \begin{subfigure}[b]{0.45\linewidth}
         \centering
         \includegraphics[width=\linewidth]{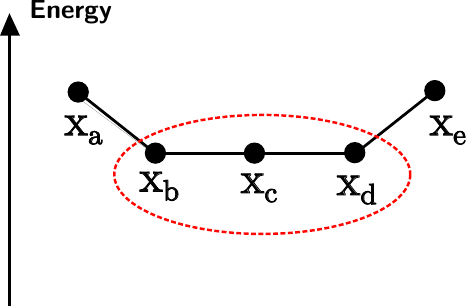}
         \caption{}
         \label{fig:plateau}
     \end{subfigure} \\[0.25cm]
     \begin{subfigure}[b]{0.9\linewidth}
         \centering
         \includegraphics[width=\linewidth]{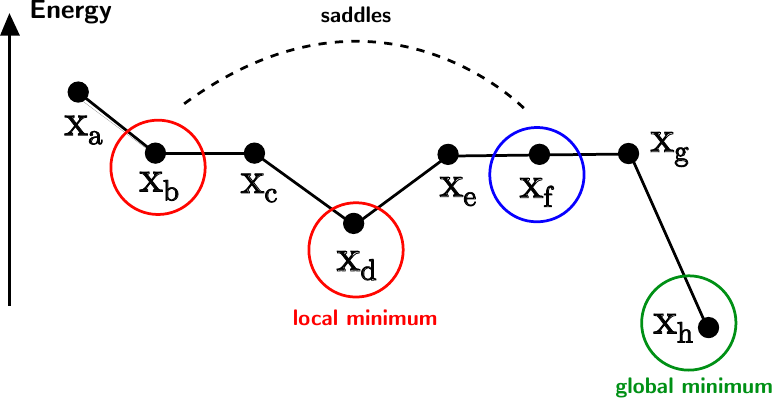}
         \caption{}
         \label{fig:saddle_connectivity}
     \end{subfigure}
    \caption{
        (a-b) Two types of degenerate landscapes: a saddle cluster and a local minimum cluster.
        State $\mathbf{x}_d$ is a zero barrier exit point from the saddle. Outlined are the stable states.
        (c) The highlighted saddle points can be treated as connected \cite{flamm2002} or disconnected
        \cite{garstecki1999} based on the adopted definition of disconnectivity graphs. Blue (red) color 
        indicates (dis)connectivity to a global minimum (green).}
    \label{fig:linear_landscape}
\end{figure}

Special attention, however, should be given 
to the  \textit{degeneracy} of such landscapes: many configurations form neighbourhoods
which can be traversed by a local rule at no energy cost. 
Furthermore, the concept of a local minimum becomes ambiguous and \textit{non-local} 
in degenerate landscapes \cite{flamm2002, hallam2005}. As Figs.~\ref{fig:shoulder} and \ref{fig:plateau}
demonstrate, it is impossible to know if a descending energy 
path exists from the leftmost stable state $\mathbf{x}_b$ unless exploration finding the rightmost 
state $\mathbf{x}_d$ is performed. In this work we will call a stable ``plateau'' of 
Fig.~\ref{fig:shoulder} \textit{a saddle cluster}, 
while the plateau in Fig.~\ref{fig:plateau} will be called \textit{a local/global minimum cluster}.
The terminology of saddles/local minima of this work is chosen to resemble similar terms from 
continuous optimization. There, multiple works highlight profound difficulties
of navigating high-dimensional landscapes arising from both types
of critical points~\cite{dauphin2014, lee2019}.

The works of \cite{flamm2002, hallam2005, biswas2020} have addressed the complexity of constructing 
DGs of degenerate landscapes with exhaustive enumeration of states. 
While being computationally infeasible for problems larger than $\approx 30\!-\!40$ variables, 
these works carried out classifications of saddles or local minima and the ways the states can be connected within 
a cluster and to other clusters.
For example, a difference in possible connectivity of stable points is illustrated in Fig.~\ref{fig:saddle_connectivity}. 
Approach of this work is closest to that of \cite{garstecki1999} in which the highlighted saddle points are 
treated as being disconnected. This choice is motivated by the golf-course-type energy landscapes of 3-XORSAT 
problems \cite{bellitti2021}, where the paths to good solutions are mostly impeded by the entropic barriers,
rather than the energy barriers. 

In Fig.~\ref{fig:saddle_connectivity} there is no barrier between $\mathbf{x}_b$ 
and the global minimum $\mathbf{x}_h$, but the path to it lies through a local minimum $\mathbf{x}_d$. 
As a result, joining the states separated by a ``hole'' would result in a deceiving visualization 
hiding landscape features important for local search routines. 
With the methods of this work (see Sec.~\ref{sec:dg_sampling}-\ref{sec:dg_bs_extension}), 
we will address such diversity of scenarios by distinguishing the states with connections to 
global minimum (blue color) from those separated from it by either barriers or ``holes'' (red color).
This will provide a clear explanation of why \textit{second} order IMs can be greatly challenged by
higher-order combinatorial optimization problems (Sec.~\ref{sec:qubo_3sat}).

\section{Results \label{sec:results}}

\subsection{Methods \label{sec:methods}}
\subsubsection{Locality reduction of k-SAT \label{sec:locality_reduction}}

The k-SAT problem of maximizing the number of satisfied clauses of Eq.~\ref{eq:k_SAT} 
is reformulated as PUBO \eqref{eq:PUBO} minimization as follows 
(inverting the expression and using De-Morgan law):
\begin{eqnarray}
    &&(l_{i_{1,1}} \vee l_{i_{1,2}} \dots \vee l_{i_{1,k}})\wedge \dots \nonumber\\ 
    &&\rightarrow \bar{l}_{i_{1,1}}\bar{l}_{i_{1,2}}\dots\bar{l}_{i_{1,k}} + \dots\,,
    \label{eq:pubo_from_cnf}
\end{eqnarray}
where each literal $l_{i} = x_{i}$ or $l_{i} = \bar{x}_{i}$.
The issue of its locality reduction to support the formulation of Eq.~\ref{eq:QUBO_H}
has been heavily investigated over the recent years \cite{dattani2019}, with 
the efforts aimed at introducing quadratizations that
have the smallest possible number of auxiliary variables,
minimize bit-precision requirements on the weights, or
have algorithmically favourable properties, e.g. submodularity \cite{schrijver2000}.
 
Perhaps the simplest method to meet the first requirement is to use quadratization by substitution, 
i.e. to introduce \textit{auxiliary} variables $y$ for each pair of variables $x_px_q$ in the original 
PUBO function of Eq.~\ref{eq:PUBO} until the problem of required order is obtained, 
i.e. 2nd order for QUBO~\eqref{eq:QUBO_H}. 
The constraints are then enforced by either explicitly considering
the equalities $x_px_q = y$, or by the addition of quadratic penalty terms in the cost 
function for each substitution:
\begin{eqnarray}
    f(\mathbf{x}) &=& \pm x_1x_2\dots x_k \to \nonumber\\ 
     &\to& g(\mathbf{x}, \mathbf{y}) =  \pm yx_3\dots x_k + P_{\pm}(x_1, x_2, y)\,.
    \label{eq:rosenberg_penalty}
\end{eqnarray}
The choice of the $P$ function is not unique; for instance, one can make sure that 
\begin{equation}
    f(\mathbf{x}) = \min_y g(\mathbf{x}, \mathbf{y})
    \label{eq:quadratization}
\end{equation}
is satisfied, thereby preserving global minima of the original problem.
Additionally, the choice of $x_i x_j$ admits some freedom and can be optimized 
for the minimum number of auxiliary variables by solving a vertex cover problem 
\cite{boros2014}. For simplicity, we use an efficient greedy 
routine to perform such optimization 
(for more details on quadratization methods outlined below cf. App.~\ref{appx:quadratization}).

For example, a commonly used quadratization penalty choice for locality reduction 
was suggested by Rosenberg (3rd order example) \cite{rosenberg1975}:
\begin{equation}
    \pm x_px_qx_k = \min_y \left[\pm yx_k  + (3y - 2x_py - 2x_qy + x_px_q)\right]\,,
    \label{eq:rosenberg_mapping}
\end{equation}
where $x_px_q$ was replaced by $y$, and the remaining terms 
penalize the mismatch of $x_p x_q$ and $y$. Thus, every appearance of $x_px_q$
in the $k\ge 3$ terms of the PUBO function is substituted 
by the same $y$, and for each such substitution $3y - 2x_py-2x_qy + x_px_q$ penalty is added.
This mapping is also implicitly used when the approach of reversible logic of \cite{biamonte2008, aadit2022} is employed.

Performing standard simulated annealing optimization of 3-SAT problems we found a different 
mapping to be computationally superior to the Rosenberg version. 
The new mapping extends the quadratization ideas \cite{kolmogorov2004, freedman2005} and \cite{boros2014} 
by approaching the monomials with positive and negative coefficients differently:
\begin{eqnarray}
    &&-x_1\dots x_k = \min_y\left[ (k-1)y - \sum_{i=1}^k x_iy \right]\,, \nonumber\\
    &&x_1\dots x_k = x_2\dots x_k -\bar{x}_1\dots x_k \nonumber\\ 
    &&= x_2\dots x_k + \min_y\left[(k-1)y - \bar{x}_1y- \sum_{i=2}^k x_iy \right]\,,
    \label{eq:KZFDBG}
\end{eqnarray}
and thus we call it KZFD-BG after the authors. However, instead of applying 
these penalties individually for each term in the PUBO function \cite{sharma2023}, 
the variable substitution in this work is done as in the Rosenberg case, i.e. sharing 
substituted pairs across multiple monomials (see App.~\ref{appx:quadratization}). 
As a result, this yields the same number of native and auxiliary variables regardless of the mapping used.
We address simulated annealing performance difference of the mappings 
in the context of PUBO and QUBO comparison in Sec.~\ref{sec:qubo_3sat}.

\begin{figure}[ht]
    \centering
    \includegraphics[width=0.95\linewidth]{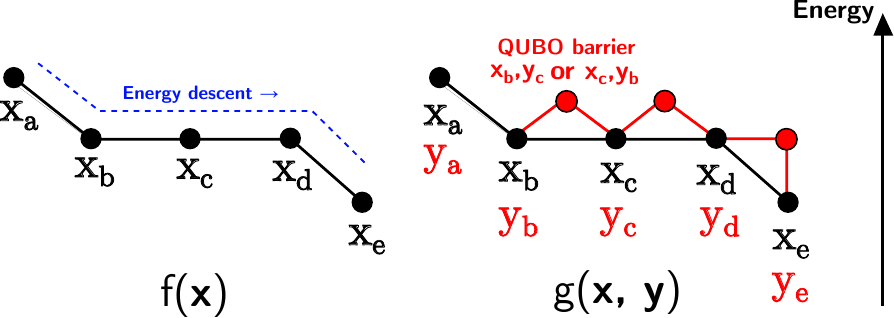}   
    \caption{(left) PUBO landscape sketch: neighbouring states are connected by a single flip.
    (right) QUBO mapping landscape; the auxiliary $\mathbf{y}$ adaptation may 
    introduce new energy barriers preventing the otherwise possible descent in energy.}
    \label{fig:pubo_v_qubo_simple}
\end{figure}

Any locality reduction method modifies the ``native'' 
optimization landscape in non-trivial ways and can make its exploration algorithmically more challenging.
In particular, Eq.~\ref{eq:quadratization} guarantees that for every stable state $\mathbf{x^*}$ of $f(\mathbf{x})$
with respect to a single bit-flip: 
$f(\dots,\bar{x}^*_i,\dots) - f(\dots,x^*_i,\dots) \ge 0,\,\forall i$,
the quadratization $g(\mathbf{x^*}, \mathbf{y^*})$ is also in a stable state, 
which is given by $\min_{\mathbf{y}}g(\mathbf{x^*}, \mathbf{y})  \equiv g(\mathbf{x^*}, \mathbf{y^*})$. 
Indeed, the bit-flip energy changes with respect to auxiliary variables are non-negative due to the definition of $g$:
$g(\mathbf{x^*}, \dots, \bar{y}_i^*, \dots) \ge g(\mathbf{x^*}, \dots, y_i^*, \dots) = \min_{y}g(\mathbf{x^*}, \mathbf{y})$. In turn, 
the energy change of flipping $x$ is also non-negative because of the following chain:
\begin{eqnarray}
    &&g(\dots,\bar{x}_i^*,\dots, \mathbf{y^*}) \ge \min_{\mathbf{y}} g(\dots, \bar{x}_i^*, \dots, \mathbf{y})\\ 
    && = f(\dots, \bar{x}_i^*, \dots) \ge f(\mathbf{x^*}) = g(\mathbf{x^*}, \mathbf{y^*})\,.
\end{eqnarray}
However, such correspondence does not hold in the opposite direction, i.e. a stable state 
of $g(\mathbf{x}, \mathbf{y})$ is not guaranteed to be a stable state of $f(\mathbf{x})$.

For illustration, in Fig.~\ref{fig:pubo_v_qubo_simple} a ``linear'' landscape represents 
states connected by a bitflip local move in the $N$ dimensional hypercube. 
The left sketch depicts a degenerate case with states $\mathbf{x}_b$,
$\mathbf{x}_c$ being stable, but $\mathbf{x}_a$ and $\mathbf{x}_d$ unstable. In turn, the right sketch
shows how the quadratization mapping induces a rugged structure on top of the original manifold due to 
the auxiliary variables and the penalty terms. For every state $\mathbf{x}$ there is a
corresponding minimizing auxiliary state $\mathbf{y}$ (possibly non-unique) 
according to Eq.~\ref{eq:quadratization}.
The low energy state $\mathbf{x}_e$ that was
easily accessible by a greedy local search descend can now be separated by energy barriers
due to the necessity to adapt $\mathbf{y}$ for every $\mathbf{x}$.

If $\mathbf{x}$ and $\mathbf{y}$ are treated on equal footing, then one is forced to 
explore a configuration space $2^{|\{y\}|}$ times bigger than the native problem.
The problem that already had highly non-trivial landscape structure caused by frustrations and
long-distance correlations of variables, after quadratization will have these features hidden or worsened by the mismatch 
of ``gradients'' and energy barriers, ultimately causing significant deterioration of the IMs' 
ability to find solutions \cite{perdomo2019, hizzani2023, sharma2023, bybee2023}.

The effect of penalty-based locality-reduction methods may be different depending on a combinatorial 
problem class that is being quadratized. For example, a popular benchmarking 
3-regular 3-XORSAT problems \cite{kowalsky2022} feature variables that appear in only three 3rd-order clauses. 
Thus, the QUBO formulation has only three native-auxiliary interactions per native variable, which are responsible 
for the QUBO energy barriers (Fig.~\ref{fig:pubo_v_qubo_simple}). 
In comparison, the phase transition random 3-SAT problems \cite{mezard2009} have on average $\approx 3\times 4.267$ 
appearances of variables in different clauses.

Finally, we note that the sparsifying approaches that aim to reduce degrees of interaction between variables 
can introduce even more energy barriers into the problem due to auxiliary variables and penalties 
akin to the locality reduction methods. We do not focus on sparsification in this work; nonetheless, one 
example is given in~App.~\ref{app:sparsification}.

\begin{figure*}[ht]
    \includegraphics[width=1\linewidth]{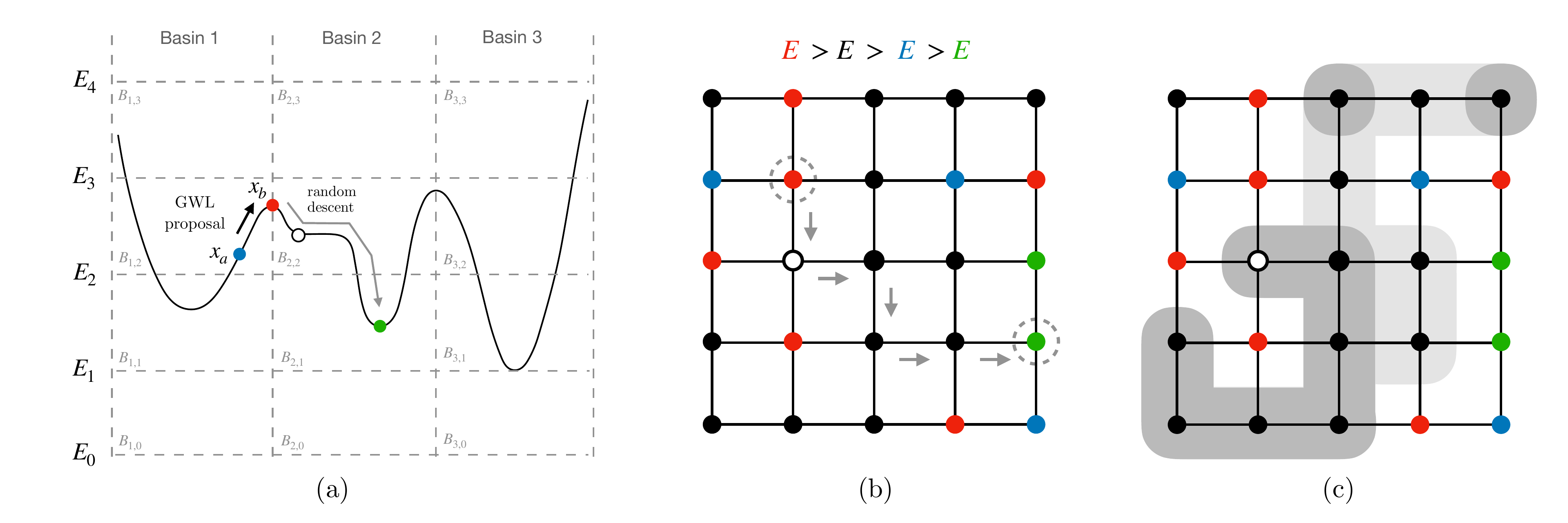}
    \caption{\label{fig:algos_summary}
        (a) Generalized Wang-Landau (GWL) for sampling and barrier estimation. GWL proposal $x_a \to x_b$ is sampled 
        with probability of Eq.~\ref{eq:GWL_prob}. The basin of attraction is identified by an algorithm of choice.
        (b) Random descent illustration. The highlighted red circle state belongs to 
        the basin of the green highlighted local minimum state.
        (c) If necessary, breadth-first search accurately calculates cluster sizes at the end of sampling. 
        The light grey states are unstable exits, the dark grey states are stable saddles.
    }
\end{figure*}

\subsubsection{Sampling algorithm outline \label{sec:dg_sampling}}
In order to study and visualize with DGs the energy landscapes of degenerate optimization problems and their QUBO 
mapping modifications, we extend the Generalized Wang-Landau (GWL) \cite{landau2004, liang2005} 
sampling approach of the works \cite{zhou2011, tang2012}. 
The GWL non-Markov Chain Monte-Carlo algorithm carries out random walks in the configuration space 
aiming to achieve approximately uniform attendance of all predefined energy levels $l \in [1, L]$ and all 
recorded basins of attraction $k \in [1, K]$.

During preprocessing steps (see Fig.~\ref{fig:algos_summary}a) one defines the landscape partition into sectors in 
energy $[E^1, E^2,\dots, E^L]$ and affinity to a basin of attraction of a local minimum $\mathbf{x}^k$: 
$\mathbf{x} \in B_{k, l}$, if $E(\mathbf{x}) \in [E^l, E^{l+1})$ 
and $\mathbf{descent}(\mathbf{x}) = \mathbf{x}^k$. We note that the $\mathbf{descent}$ routine can be defined differently, 
and it makes sense to choose its definition similar to the the actual solver algorithm that would be used for 
solving studied problems in practice (see below Sec.~\ref{sec:dg_bs_extension}).
Next, the sampling of states is performed
with the following acceptance probability:
\begin{equation}
    p_{a\to b} = \min{\left[1, \exp{\left\{\beta (E_a - E_b)\right\} \frac{\gamma_{k_a, l_a}}{\gamma_{k_b, l_b}}}\right]}\,,
    \label{eq:GWL_prob}
\end{equation}
where $\gamma_{k, l}$ is a current estimation of the statistical weight of a sector:
\begin{equation}
    \frac{1}{Z}\sum_{\mathbf{x} \in  B_{l, k}}\exp{(-\beta E(\mathbf{x}))} \approx \gamma_{k,l}\,.
    \label{eq:gamma_def}
\end{equation}
This estimation is constantly updated, when the sector $B_{l, k}$ is visited, by:
\begin{equation}
    \gamma_{l, k}^{t+1} = \gamma_{l, k}^{t}e^f\,,
    \label{eq:f_def}
\end{equation}
where $f$ can follow a decreasing schedule usually starting from the value $f = 1$.
It is numerically convenient to also define a histogram 
\begin{equation}
    \theta^{t+1}_{l, k} \equiv \ln\gamma^{t+1}_{l, k} = \theta^{t}_{l, k} + f\,, 
    \label{eq:gwl_histogram}
\end{equation}
which is initialized at 0 for all $l,k$ at the beginning of the algorithm.
If the exploration of as many minima as possible is preferred, then $f$ is not decreased 
over time \cite{tang2012}, but in this case the estimation of $\gamma$ would not be accurate \cite{barbu2020}.
We do not decrease $f$, because our goal is construction of DGs with rapid discovery of distinct local minima/saddles.

When a step $\mathbf{x}_a \to \mathbf{x}_b$ is tried (accepted or not), one saves 
the energy $\max{[E_a, E_b]}$ ($\max{[E_a, E_a + \Delta E_{\rm{QUBO}}(\mathbf{x}_a \to \mathbf{x}_b)]}$ for QUBO, see Sec.~\ref{sec:dg_qubo_extension}) 
as a current energy barrier estimation between basins $k_a$ and $k_b$.
This ``educated guess'' can then potentially be improved with the ridge descent algorithm \cite{zhou2011}.
If the zero energy barrier is found for a state perceived as local minimum
(e.g.~$\mathbf{x}_f \to \mathbf{x}_g \to \mathbf{x}_h$ in Fig.~\ref{fig:saddle_connectivity}), 
the status of such minimum is changed to a saddle, and its histogram is joined (max values of each row) with the corresponding
lower basin of attraction (e.g. $\mathbf{x}_h$). If a saddle is connected to several lower basins, 
then the visits are distributed uniformly at random among them.

We keep track of maximum $K$ number of lowest in energy local minimum/saddle clusters adaptively
uniting them by discovered connectivity and thus allowing space for additional clusters to be taken into account. 
If $K$ is too small, then only a few energy levels will be available for the DG construction.
In addition, we define a special $K+1$ column of the histogram for all of the states that do not fit into 
the first $K$ clusters \cite{tang2012}.

The implementation of the sampling method of this work is publicly avaiable for reproducibility of the results 
and is described in App.~\ref{appx:code_availability}.
Additional details, including uniformity of sampled histograms, 
accuracy of DG construction, computational cost, and hyperparameters
for all disconnectivity graphs of this work are presented in 
App.~\ref{appx:sampling_uniformity}, \ref{appx:dg_convergence}, \ref{appx:sampling_complexity}.

\subsubsection{Extension for degeneracy \label{sec:dg_bs_extension}}
By design, GWL uniformly samples states across basins of attraction of local minima and energy levels. 
Its main purpose in this work is to discover as many regions of the landscape as possible without 
being stuck in a particular place, thereby not biasing the DG estimation.
What is crucial in the definition of the algorithm is the $\mathbf{descent}$ routine, which identifies 
local minima and saddle points. When a problem has no degeneracies, e.g. S-K spin glass with Gaussian 
weights, one can define the descend (hill climbing) as the steepest descent, i.e. 
spins with the highest energy reduction are flipped. However, in this work we are interested in highly degenerate 
integer valued optimization problems, where such definition is not possible.

As briefly discussed above in Fig.~\ref{fig:linear_landscape}, local minima and  
saddles are perceived differently depending on the algorithm employed for solving such problems. 
In this case, constructing exact DGs, apart from being infeasible for large problems, 
may result in misleading conclusions. For instance, 
a very large saddle point may have only one zero barrier exit from itself, 
which may never be found by a local 
search routine, effectively being a local minimum, but it would still be depicted as a saddle on a DG, 
or even worse, not shown at all if saddles are not considered.

Here we aim to balance between efficient exploration of the landscape and visualization of relevant 
landscape features. For this purpose, we use \textit{random descent} (see Fig.~\ref{fig:algos_summary}b),
in which a greedy local move is performed in the first-seen random direction decreasing the energy. 
This descent routine corresponds to the MC sampling approach we use in simulated annealing benchmarking but at $T=0$ 
(see App.~\ref{appx:sa_details} for more details on SA). 

Once a stable state is encountered (white circle in Fig.~\ref{fig:algos_summary}b), a limited exploration of the 
``plateau'' region is performed until either the budget of allowed moves is exhausted, or an exit from the saddle 
is found (green circle in Fig.~\ref{fig:algos_summary}b). 
We defined a hyperparameter which determines for how long an algorithm can explore a stable cluster
before registering it as a local minimum/saddle in the histogram. If a cluster is easily escapable,
there is no reason to keep track of it. 

The states encountered during such exploration 
of saddles/local minima are stored in a single cluster (including the unstable exit states, i.e. 4 states 
are stored in Fig.~\ref{fig:algos_summary}b). If some of the stored states are encountered again during GWL sampling,
all of the states that belong to a single cluster are joined, with their histograms united by their max values.

Previously, the works \cite{hartmann2000, mann2010} addressed the difficulty of clustering 
in the context of improving the uniform sampling of the \textit{ground states}. Once a ground state $\mathbf{x}$ 
was found, a ballistic search (BS) routine was carried out: starting from some global minimum state,  
a chain of zero energy states was constructed by flipping every variable maximum once. 
With the use of such chains, the cluster sizes and thus connectivity of states were estimated more reliably. 
We experimented with this method for clusters at every energy level and found it useful for clustering 
remote configurations when the number of states becomes infeasibly large.

Finally, in Fig.~\ref{fig:algos_summary} we illustrate breadth-first search (BFS) that we use 
to exactly evaluate sizes of clusters at the end of sampling, when such statistics are of interest.
Both stable and unstable states (exits) participate in BFS, and we confine their number by a predefined 
bound of states per energy level (usually $10^7$ in this work). While only the stable states are later shown 
on the DGs, the ratio of stable to unstable number of configurations can potentially be used to estimate the 
probability of escaping saddle regions of the landscape.

\subsubsection{Extension for QUBO mapping\label{sec:dg_qubo_extension}}

Two neighbouring configurations are considered to be a part of a single degenerate cluster in 
the native (PUBO Eq.~\ref{eq:PUBO}) landscape 
if a local move separating them is of zero energy cost, as shown for states $\mathbf{x}_b$
and $\mathbf{x}_c$ in Fig.~\ref{fig:pubo_v_qubo_simple}. 
The state $\mathbf{x}_d$ does not belong to a cluster 
since it has a move of negative energy to the state $\mathbf{x}_e$. However, in the special 
case of the QUBO mapping landscape, the same state $\mathbf{x}_d$
would now be considered a saddle point since the decrease in energy is only achieved through an intermediate
$\mathbf{y}$ adjustment.

The presence of a barrier in QUBO for a transition 
$\mathbf{x}_b \to \mathbf{x}_c$ (when originally there could be no barrier at all) puts the local search 
at a disadvantage due to the higher rejection rate of local moves. Raising the temperature of sampling, e.g. of 
simulated annealing, would not fully solve the problem since it would harm the necessary exploitation of the low-energy manifold.
Additionally, once local search is complete, a solver discards $\mathbf{y}$ values using the states of $\mathbf{x}$ as a solution.
The search over the subspace of $\mathbf{y}$, thus, does not look for new solutions, but rather varies the induced QUBO barriers 
between the neighbours in the $\mathbf{x}$ space.

\begin{figure}[ht]
    \centering
    \includegraphics[width=1\linewidth]{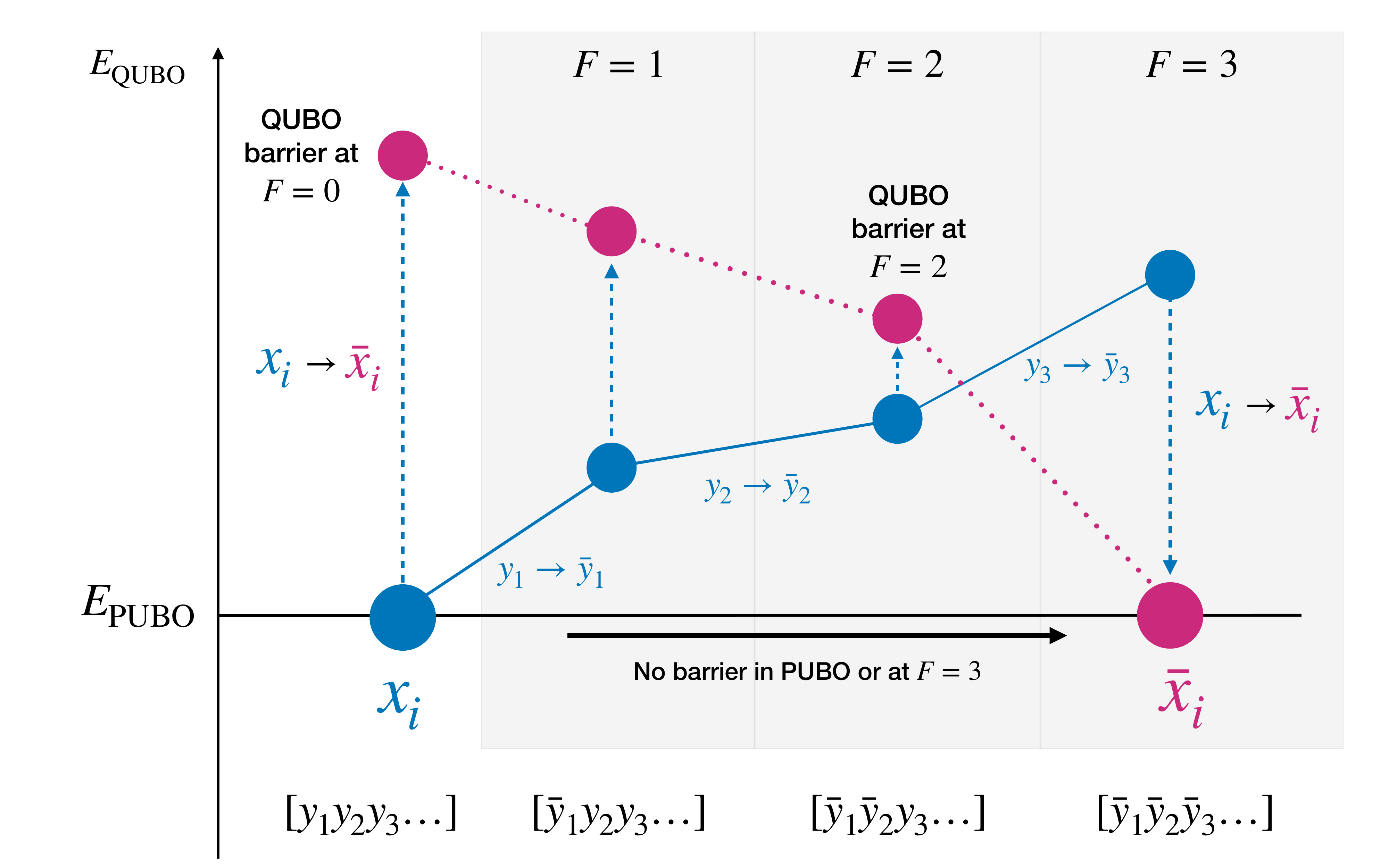}
    \caption{
        QUBO factor $F = |\{y_a\}|$ motivation example. 
        By perturbing $F = 3$ auxiliary variables 
        one is able to restore the PUBO zero barrier between $x_i$ and $\bar{x}_i$ states.
        For $F \in (0, 3)$ the effective barrier is defined, 
        taking intermediate values between QUBO and PUBO.
    }
    \label{fig:qubo_factor}
\end{figure}

To highlight the significance of landscape ruggedness of quadratization compared to the native space
and facilitate fairer comparison, here we provide QUBO with additional capabilities by assuming that
a local search solver can ``look beyond'' the QUBO landscape barriers to a certain adjustable degree.
Fig.~\ref{fig:qubo_factor} depicts the case where the penalty terms of a QUBO mapping introduce 
interactions that favour auxiliary variable states different by Hamming distance 3 
for two native configurations separated by a single bit-flip, i.e. $x_i$ and $\bar{x}_i$. 

If the problem is approached head-on, one would need to either climb a steep barrier of $x_i\to \bar{x}_i$
and then adapt 3 auxiliary variables, or sequentially flip each of $y_a \to \bar{y}_a$, i.e. climb a long barrier.
Such scenario of long barriers is argued to be difficult for tunnelling in quantum annealers \cite{denchev2016}, 
considering that the mapping quadratization is essential due to strict hardware limitations.
We note, however, that with every sequential flip of $y_a$, the barrier 
of the $x_i\to \bar{x}_i$ move is reduced, i.e. allowing more $y_a$ to be explored
raises the chance to overcome the QUBO barriers introduced by the mapping in the first place.

As a result, we augment disconnectivity graph analysis by introducing a QUBO factor $F$, which
stands for the maximum number of allowed auxiliary variable flips of non-zero energy 
for every native move $x_i\to \bar{x}_i$.
With large enough $F$ the original (PUBO) landscape is recovered, while for small $F$ values 
``effective'' energy barriers are still present, and thus the landscape connectivity 
is worsened by the mapping. In addition, $F$ serves as means to compare different QUBO mappings head-to-head,
with mappings allowing small $F$ being arguably better for the local search of IMs. 
We perform such comparison supported by the simulated annealing results in Sec.~\ref{sec:qubo_3sat}.

The algorithm to compute the effective barriers is as follows. 
First, at a fixed position in the space of $\mathbf{x}$ 
we set the auxiliary variables $\mathbf{y}$ in a valid state required by Eq.~\ref{eq:quadratization} \footnote{
    This valid state may not be unique, i.e. there could be degeneracy in the auxiliary variable space $\mathbf{y}$ 
    for each fixed $\mathbf{x}$. In this case we consider all such configurations and take the minimum 
    of the computed effective QUBO barrier.
}.
Next, the bit-flip energy change $\Delta E_{\mathrm{QUBO}}(\dots, x_i \to \bar{x}_i, \dots \mathbf{y})$ is computed, which 
corresponds to the ``vanilla'' QUBO barrier at $F = 0$. 
Second, in order to calculate the effective QUBO barrier of $x_i \to \bar{x}_i$, we list all auxiliary variables $\{y_a\}$
that interact with $x_i$, i.e. $Q_{ia} \ne 0$.
Out of all listed $y_a$, we choose 
$F$ variables with the minimum values of $\Delta E(\dots,\bar{x}_i,\dots, y_a \to \bar{y}_a,\dots) < 0$. 
Finally, the effective barrier (see Fig.~\ref{fig:qubo_factor}) $\Delta E_{\mathrm{QUBO, F}}$ is obtained by 
($y$ variables don't interact with each other in 3-SAT mappings):
\begin{equation}
    \begin{aligned}
    &\Delta E_{\mathrm{PUBO}}(x_i \to \bar{x}_i) \le \Delta E_{\mathrm{QUBO, F}}(x_i \to \bar{x}_i) \equiv\\
    &\Delta E_{\mathrm{QUBO}}(x_i \to \bar{x}_i) + \sum_{a=1}^F \Delta E(\bar{x}_i, y_a \to \bar{y}_a) < \\
    &\Delta E_{\mathrm{QUBO}}(x_i \to \bar{x}_i)\,.
    \label{eq:effective_barrier}
    \end{aligned}
\end{equation}

\subsubsection{Disconnectivity graphs notation \label{sec:dg_notation}}
In the following sections we adopt the following convention when plotting DGs 
(e.g. see Fig.~\ref{fig:easy_and_hard_dg}). The y-axis stands for 
the PUBO/QUBO energy. Every circle represents a separate local minimum/saddle cluster. 
The diameter of such circle corresponds to the square root of the cluster
degeneracy, i.e. the area of a circle is proportional the number of connected stable configurations within a cluster. 
There is no explicit meaning behind the x-axis distance between the DG leaves and branches. 
If a circle is shown to have a zero-energy connection to lower clusters, then it represents a saddle cluster.
If two or more saddles appear connected, then
the situation depicted in Fig.~\ref{fig:saddle_connectivity} 
between $\mathbf{x}_b$ and $\mathbf{x}_f$ is in place. 
Red clusters in Figs.~\ref{fig:DG_uf_vs_semi}a and ~\ref{fig:DG_uf_vs_semi}b 
have no direct connection (not found during sampling) to the global minimum denoted by green, 
i.e. all local minima are red, as well as some saddles (e.g. the state 
$\mathbf{x}_b$ in Fig.~\ref{fig:saddle_connectivity}). 
Blue saddle clusters in Figs.~\ref{fig:DG_uf_vs_semi}a and ~\ref{fig:DG_uf_vs_semi}b 
were found to be connected to the global minimum by a descent algorithm of choice 
without energy barriers (e.g. $\mathbf{x}_f$ in Fig.~\ref{fig:saddle_connectivity}).

Every DG is accompanied by a histogram of the number of states obtained with BFS 
at each energy level. The degeneracy of every separate cluster is denoted by $\mathcal{N}_k$, while 
the total number of states per energy is plotted as a normalized by $N$ (number of native variables) 
natural logarithm of $\sum_k \mathcal{N}_k$. The grey histogram shows the total number of BFS aggregated 
states (including unstable saddle exits). The blue and red histograms count the corresponding stable states shown 
by circles on the DG.

\subsection{Easy and hard problems \label{sec:easy_and_hard_dg}}

\begin{figure}[h!]
    \centering
    \begin{subfigure}[b]{1\linewidth}
        \centering
        \includegraphics[width=\linewidth]{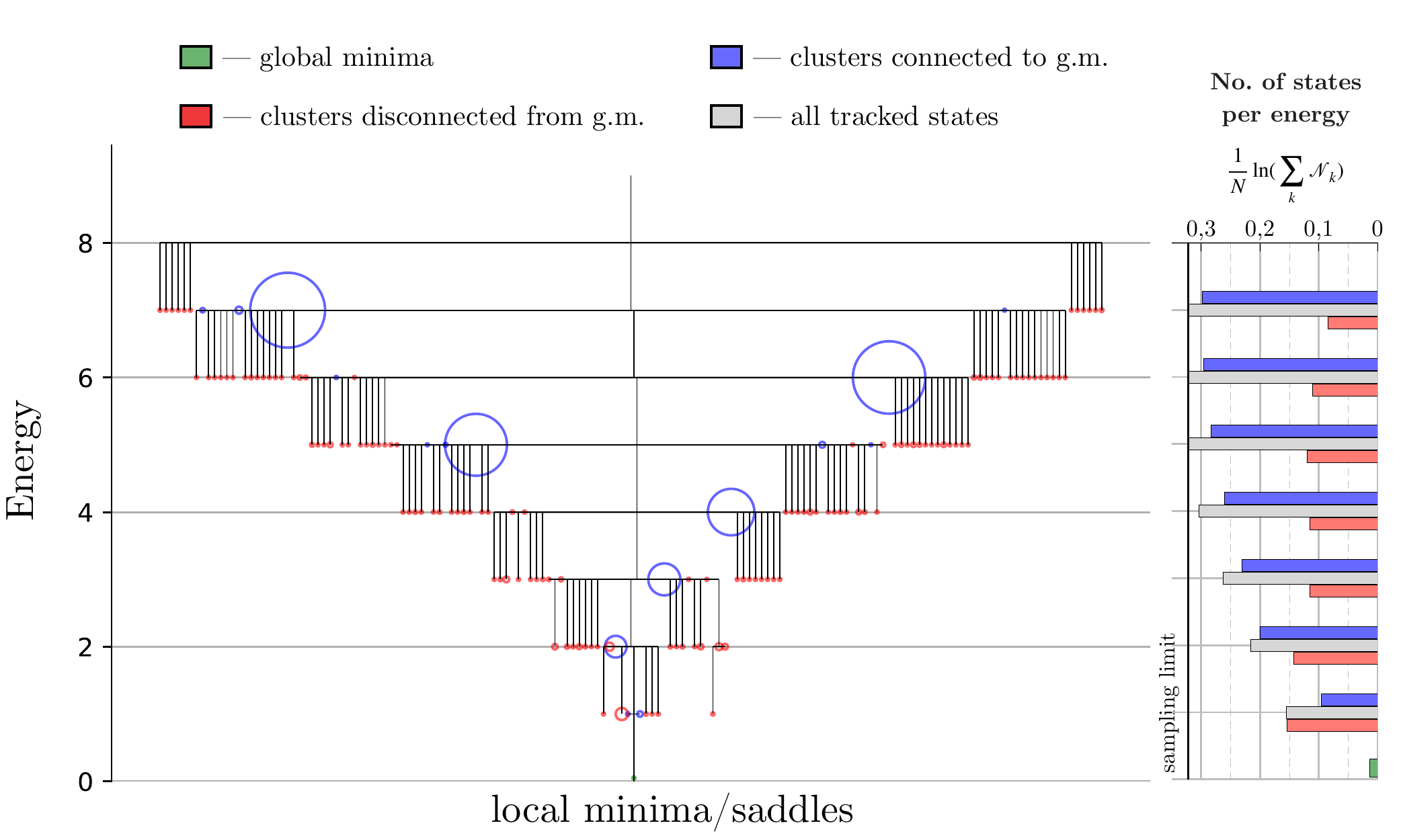}
        \caption{}
        \label{fig:map0_uf981}
    \end{subfigure}
    \hspace{0.03\linewidth}
    \begin{subfigure}[b]{1\linewidth}
        \centering
        \includegraphics[width=\linewidth]{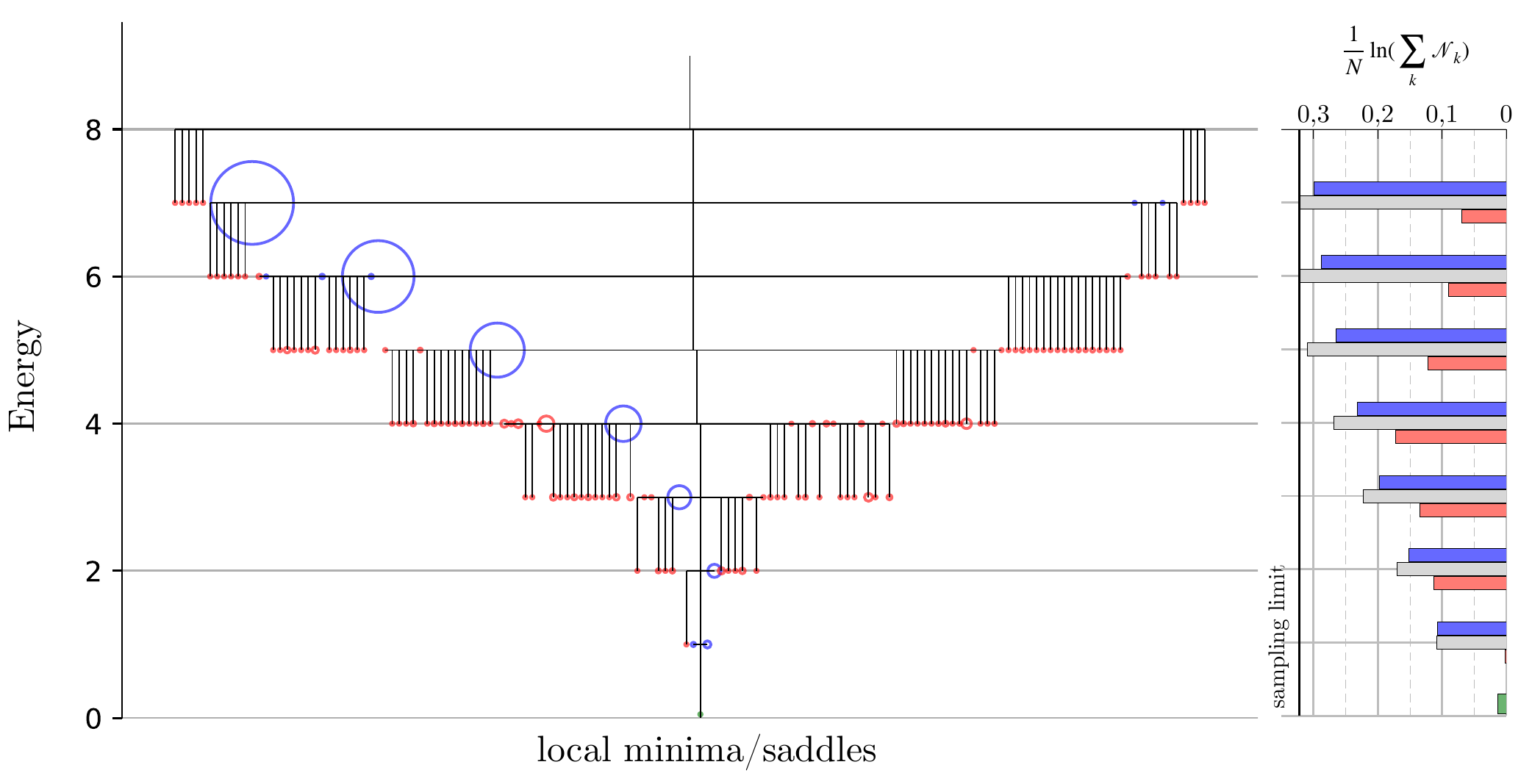}
        \caption{}
        \label{fig:map0_uf920}
    \end{subfigure}
    \hspace{0.03\linewidth}
    \begin{subfigure}[b]{1\linewidth}
        \centering
        \includegraphics[width=\linewidth]{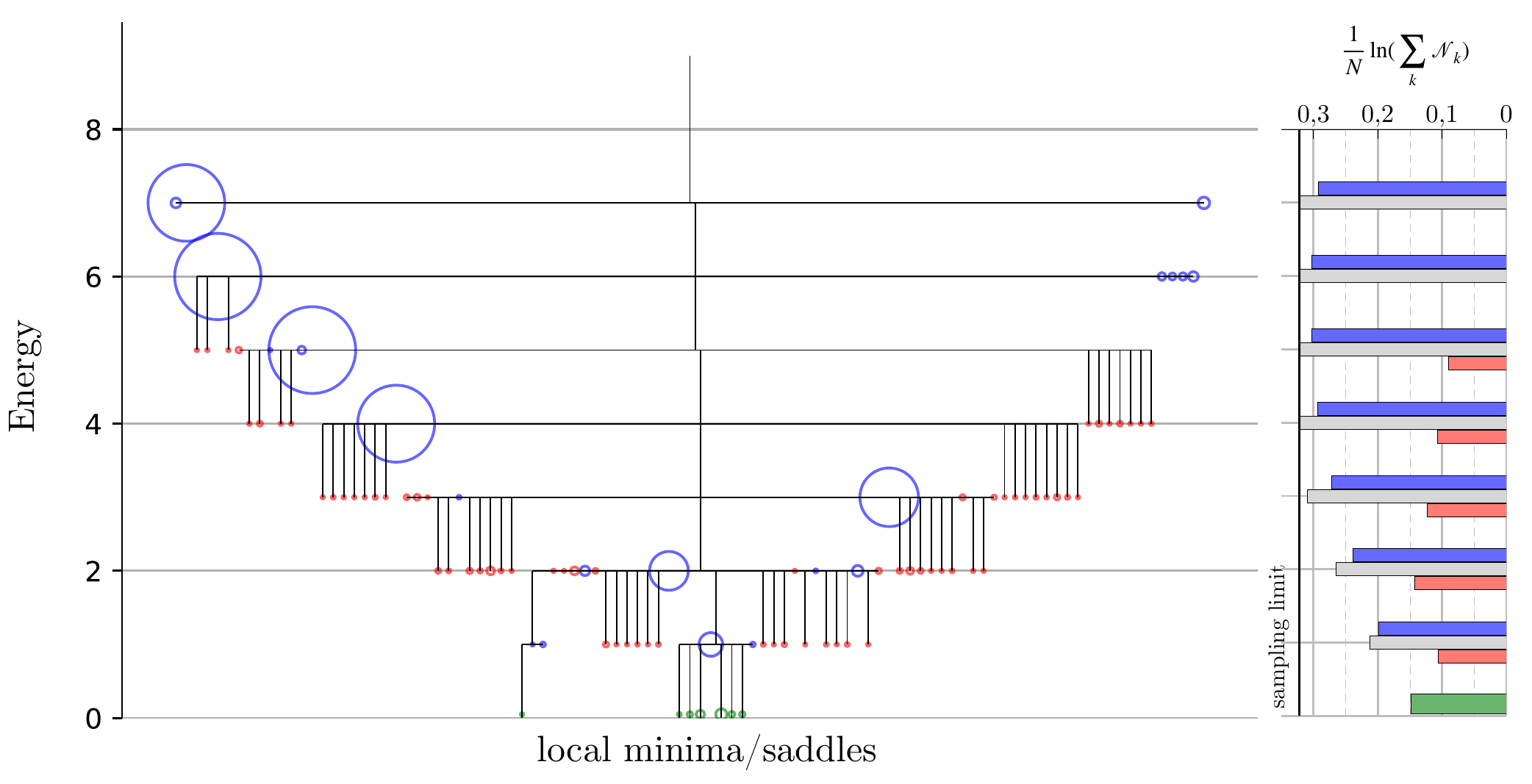}
        \caption{}
        \label{fig:map0_uf933}
    \end{subfigure}
    \caption{
        Disconnectivity graphs (PUBO) of ``very hard'' (a, instance \textit{uf50-981}),
        ``hard'' (b, instance \textit{uf50-920}),
        and ``easy'' (c, instance \textit{uf50-933}) 3-SAT problems. 
        States truncated at $E \le 7$. Sampling limit per energy level: $10^7$. 
        (a) 156 clusters, 2 global minimum states.
        (b) 148 clusters, 2 global minimum states.
        (c) 99 clusters, 1654 global minimum states.
    \label{fig:easy_and_hard_dg}}
\end{figure}

\begin{figure}[ht]
    \centering
    \includegraphics[width=1\linewidth]{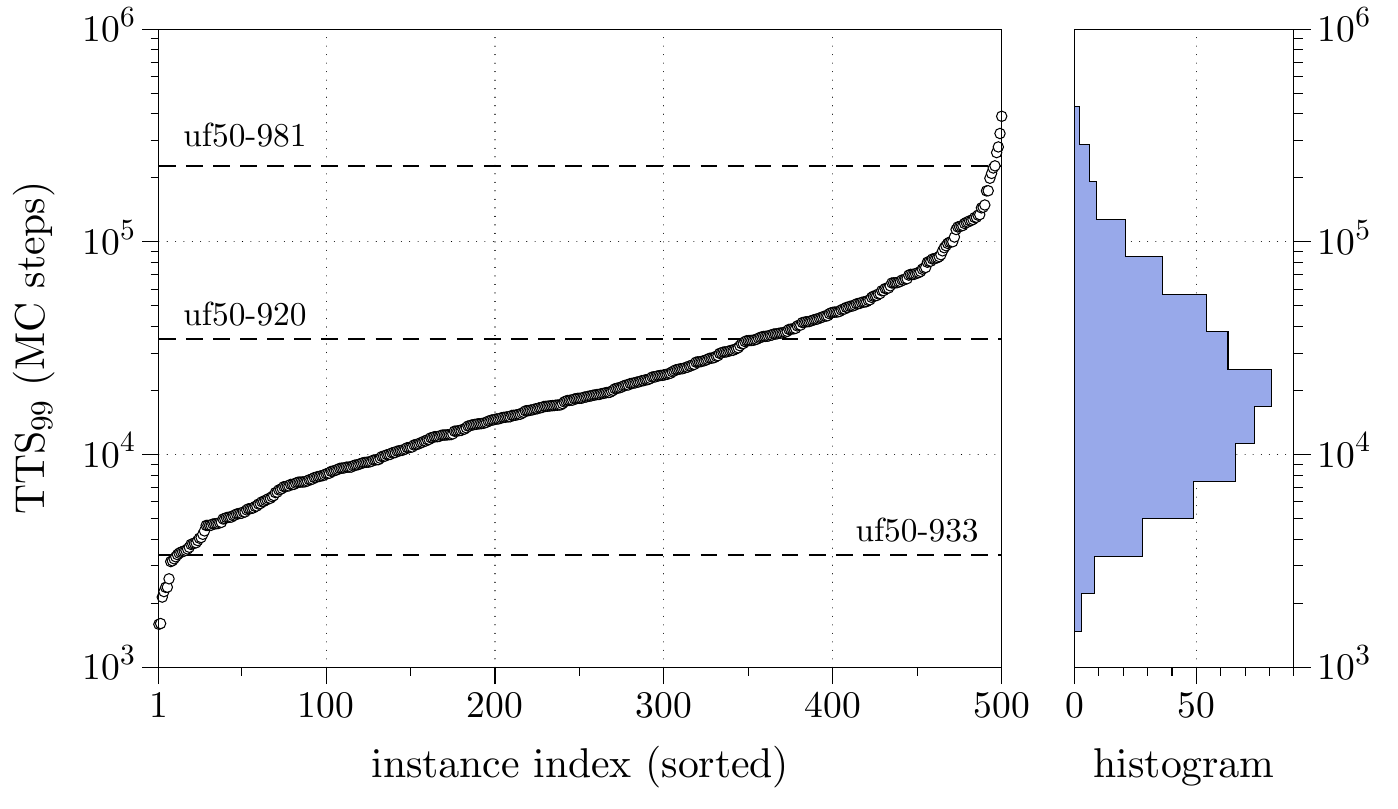}
    \caption{The distribution of the TTS$_{99}$ of the PUBO (native)
    SA for 500 instances (\textit{uf50 500-1000}). 
    Dashed lines for the TTS$_{99}$ of instances used for DG construction in Fig.~\ref{fig:map0_uf920}.
    For SA implementation details and code availability cf.~App.~\ref{appx:sa_details}.}
    \label{fig:pubo_50_tts_hist}
\end{figure}

The finite size fluctuations of relatively small random 3-SAT problems usually employed for 
IM benchmarking in practice results in a strong spread of their hardness \cite{hizzani2023}. 
In this section, with the help of DGs and using the open benchmarking library SATLIB \cite{satlib}, 
we aim to highlight the landscape features exhibited by such instances of different hardness.
As means of benchmarking we employ a simulated annealing (SA) solver described in detail 
in App.~\ref{appx:sa_details} and available at \cite{sat-qubo-dg-sampler}. 
For every optimization problem instance it outputs
time-to-solution $99\%$ (TTS$_{99}$) value (in Monte-Carlo steps), which is the time 
needed for a stochastic solver to reach a solution at least once with probability $p=0.99$. 

In Fig.~\ref{fig:pubo_50_tts_hist} we show the SA hardness distribution of 500 instances from SATLIB of size 
$N=50$ and $M=218$ clauses ($\alpha = M/N$ near the phase transition).
The instance \textit{uf50-920} visualized by the DG in
Fig.~\ref{fig:map0_uf920} was found to be relatively
``hard'' with $(34.9 \pm 1.0)\times 10^3$ TTS$_{99}$ algorithmic steps,
the instance \textit{uf50-933} in Fig.~\ref{fig:map0_uf933} was 
``easy'' with~$3390 \pm 40$ steps, and
the instance \textit{uf50-981} in Fig.~\ref{fig:map0_uf981} was 
``very hard'' with~$(229 \pm 16)\times 10^3$ steps (with respect to the observed range of TTS).

A clear distinction is seen in both the number of global minimum configurations, 
as well as the number of distinct global minimum clusters between the easy and hard instances (7 vs 1).
It is in general unclear what is the property of optimization landscapes that would 
measure hardness best for a particular solver. In \cite{mohseni2021} authors use the number of global  
minimum clusters as a proxy for predicting Survey Propagation's ability to find global minima in 4-SAT 
random instances.
However, instead of trying to predict the hardness of instances by DGs,
we demonstrate how DGs can be used to gain insights into experimentally observed 
algorithm behaviours by a diverse set of sampled landscape properties.

In addition to much larger cardinality of the set of global minima, the easy problem in Fig.~\ref{fig:map0_uf933} features 
many saddle point states at $E = 1$ ($20877$ states) which are only connected to global minima  
and act as a basin of attraction for solvers. In comparison, the ``hard'' instance 
contains similar saddles with only $206$ states. There were no local minima found 
at $E > 5$ in the easy instance, while the ``hard'' and ``very hard'' instances feature local minima even at $E = 8$.
We also highlight the higher ratio of local minima/disconnected saddles in the 
``hard'' instance compared to the ``easy'' one (the number of ``red'' states).

At every energy level we observe massive saddle points which are connected to the global minimum.
We note, however, that it becomes very important for local search not to descend into a wrong local minimum cluster, 
even though in principle it is possible to descend to a solution without overcoming any barriers. 
This is particularly highlighted by the difference between ``hard'' and ``very hard'' instances. 
In the histogram of Fig.~\ref{fig:map0_uf981} we observe a large number of local minima at $E=1$, as well as a 
distinct basin of attraction separated from the global minimum by the barrier $\Delta E=2$, 
compared to Fig.~\ref{fig:map0_uf920}.

The majority of energy barriers in the tested 3-SAT problems is the minimum possible one, $\Delta E = 1$ 
(as also previously observed in \cite{frank1997}).
In other words, the constructed DGs illustrate the significance of entropic barriers
that are determined by probabilities of descending into better areas of the landscape, which resulted in 
the observed more than two orders of magnitude spread of the time-to-solution metric in Fig.~\ref{fig:pubo_50_tts_hist}.

\subsection{Random and industrial problems \label{sec:rand_industrial_dg}}

The uniform random 3-SAT problems are a common benchmark for testing the 
performance of heuristic solvers. 
In the thermodynamic limit of $N \to \infty$ and $M \to \infty$ 
their static properties are understood within the framework of the replica 
symmetry breaking (or cavity) methods of statistics \cite{montanari2008}.
In general, the lack of structure of random CSP causes state-of-the-art exact solvers 
to struggle near the phase transition ratio 
$\alpha = M/N$ and ultimately take exponential time to find solutions due to the 
difficulty of truncating the search space based on exponentially growing deep decision 
trees \cite{selman1996}. 

On the other hand, it is not a difficult task to engineer a structured problem to
challenge a heuristic solver. A very small basin of attraction
of a global minimum with overall rugged landscape would make a local search heuristic
relying on stochastic exploration get lost. 
As a result, stochastic by design, IMs can have a hard time outperforming 
exact routines exploiting inherent structures of problems. In order to draw conclusions 
about the capabilities that IMs would need to tackle both combinatorial 
optimization classes, we employ DGs to visualize the distinction 
in landscape properties between fully random and structured ``industrial'' instances.

\begin{figure}[ht]
    \begin{subfigure}[b]{1\linewidth}
         \centering
         \includegraphics[width=\linewidth]{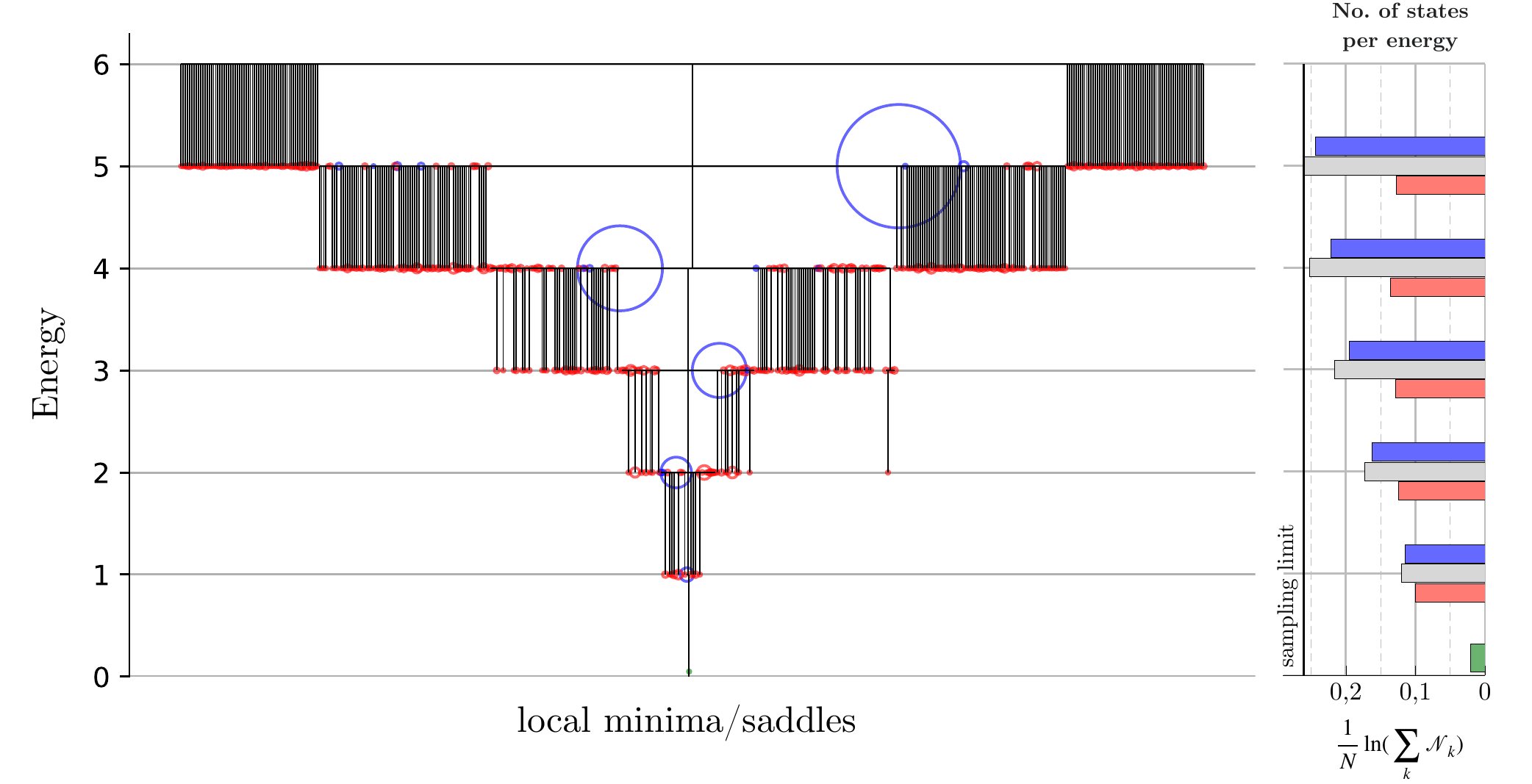}
         \caption{}
         \label{fig:map0_n68uf1}
     \end{subfigure}
     \hspace{0.0\linewidth}
     \begin{subfigure}[b]{1\linewidth}
         \centering
         \includegraphics[width=\linewidth]{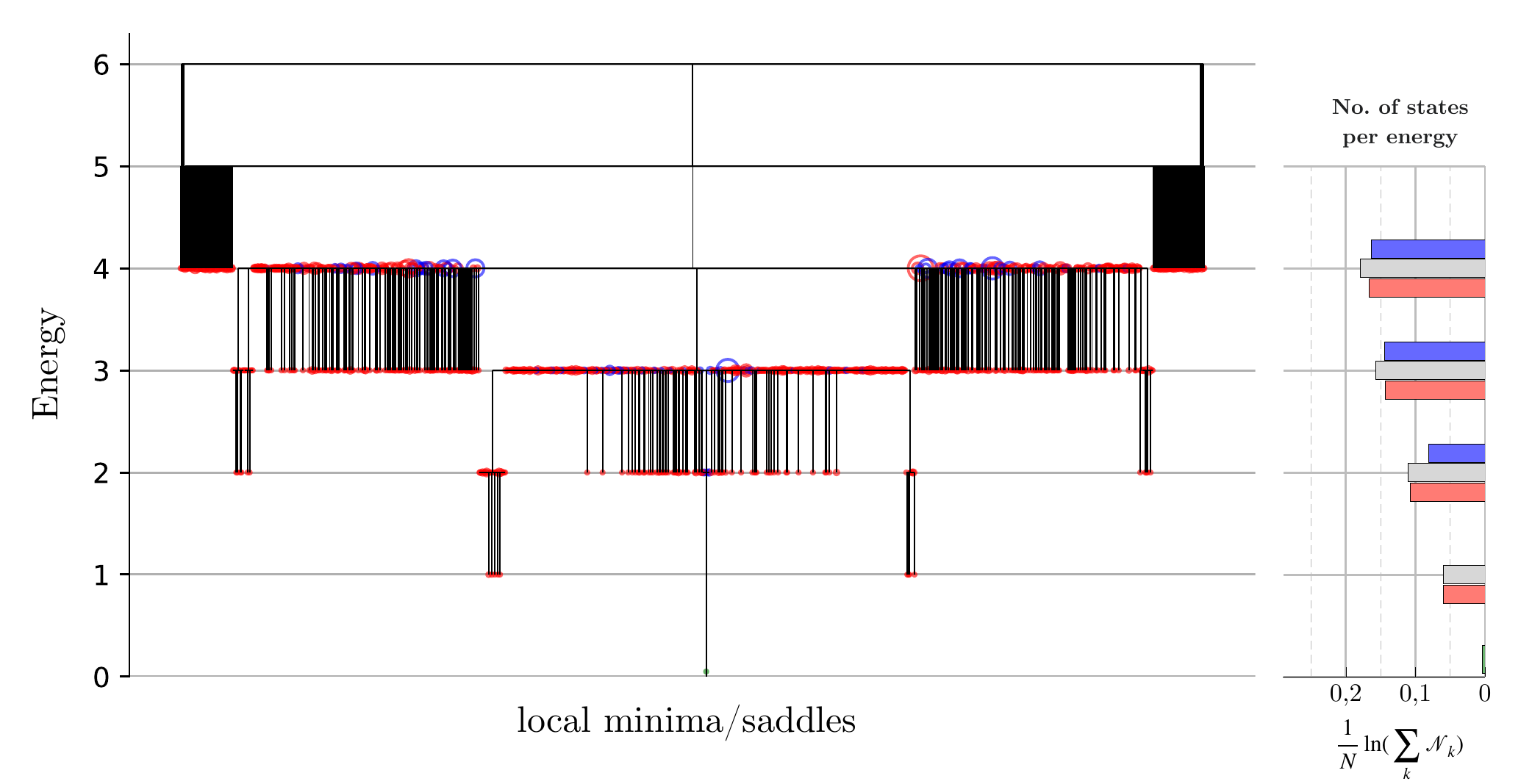}
         \caption{}
         \label{fig:map0_n68semi1}    
     \end{subfigure}
     \begin{subfigure}[b]{0.494\linewidth}
         \centering
         \includegraphics[width=\linewidth]{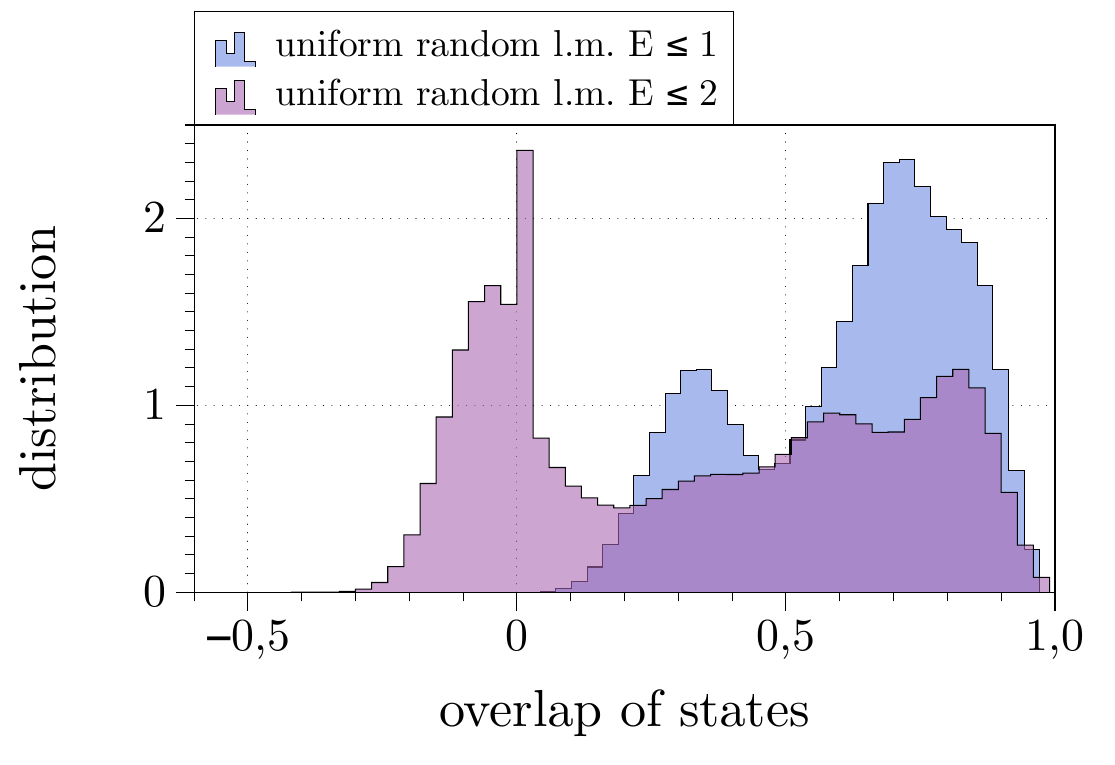}
         \caption{}
         \label{fig:uf_overlap}    
     \end{subfigure}
     \begin{subfigure}[b]{0.494\linewidth}
         \centering
         \includegraphics[width=\linewidth]{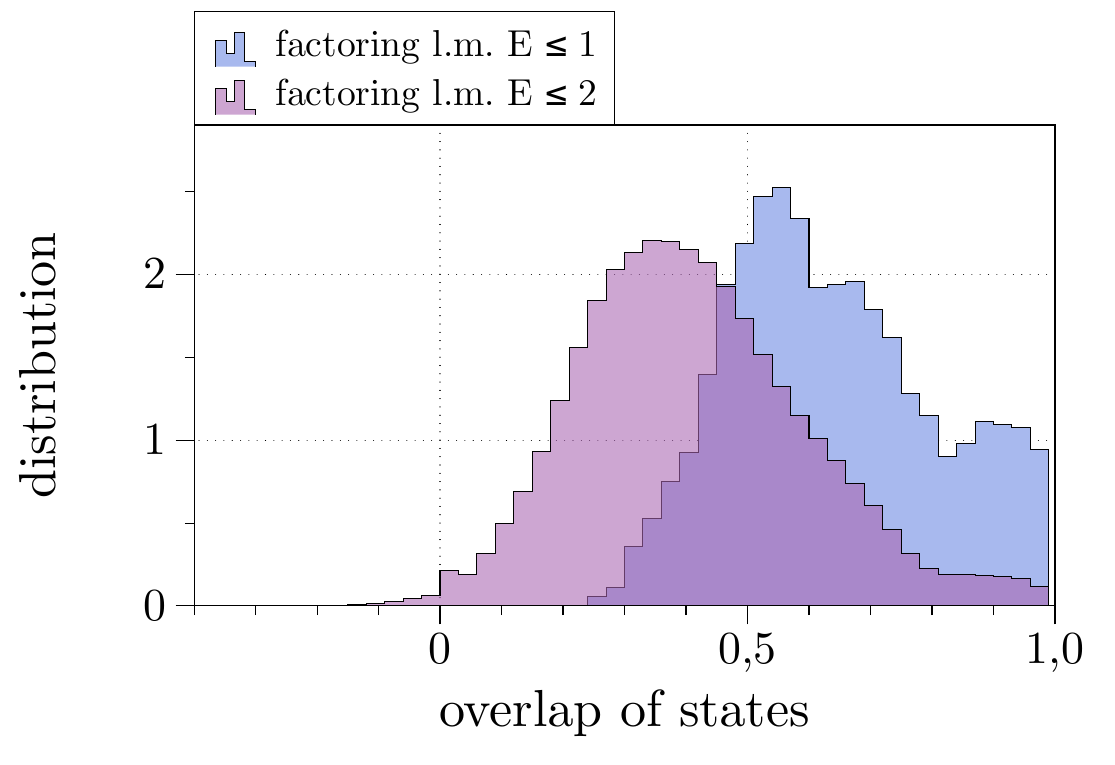}
         \caption{}
         \label{fig:semi_overlap}    
     \end{subfigure}
    \caption{Disconnectivity graph examples of 3-SAT problems with local minima truncated at $E \le 4$:
        (a)~Uniform random of 68 variables and 295 clauses. 
        $5\times 10^7$ sampling limit of states per energy. 474 clusters, 4~global minima.
        (b)~Semi-prime factoring of 55 mapped to the 3-SAT problem of 68 variables and 248 clauses \cite{semiprime}.
        998 clusters, 1~global minimum.
        (c-d)~Overlap distributions of local minimum states.
    }
    \label{fig:DG_uf_vs_semi}
\end{figure}

To represent the structured industrial class, we generated a 3-SAT formulation of the 
factoring problem of the number 55 using the method from \cite{semiprime}. 
This resulted in a 3-SAT instance having 68 boolean variables and 248 clauses
with only one global minimum. For comparison, a random uniform instance 
near the phase transition ratio $\alpha$ was generated of the same 68 variable size, 
but with 295 clauses. We obtained an instance with a single global minimum 
cluster having 4 configurations. The DG of the uniform random
instance is shown in Fig.~\ref{fig:map0_n68uf1}, 
while the DG of factoring --- in Fig.~\ref{fig:map0_n68semi1}.

With the chosen value of $K = 500$, the DG of the random problem 
was truncated at the energy levels $E = 5$ or lower, resulting in $474$ 
distinct local minimum/saddle clusters after post-processing. 
In comparison, the semi-prime factoring DG has managed to fit only clusters 
at the energies $E \le 4$ with $ K = 1000$.
This indicates much more pronounced ruggedness of the factoring 3-SAT problem 
with weak connectivity of saddle points.

As in the previous section, we observed exponentially large connected saddle clusters at every energy level the 
of random 3-SAT instance:
it is possible to traverse huge distances in the optimization landscape without the need
to overcome any energy barriers. It means that the 
hardness of this problem class arises mainly from the entropic barriers, 
leaving gradient-based solvers oblivious about 
meaningful exploration directions. 

In comparison, the number of saddle clusters disconnected from the global minimum 
in the factoring problem constitute a much bigger fraction of the overall number of  captured clusters.
Moreover, the number of stable states of the factoring problem at the given energies is much
smaller than in the random case (we didn't need to impose sampling limits), 
indicating that the energy barriers are the main contributor to 
the hardness. These results support the conclusion of \cite{mosca2022}, where authors argue that there is 
no evidence for an advantage of employing SAT reductions for factoring problems, 
both using classical SOTA SAT solvers, and their hypothetical classical or quantum 
physics-inspired counterparts.

The limitation of DGs is that they compress combinatorial landscape information 
to local minima and barriers between them, while the distances in solution space 
are left aside. Since we are able to store all of the discovered by GWL+BFS states, 
one approach to probing such distance information is by calculating the mutual overlaps of local minima.
The mutual overlap of states for Ising formulated problems is defined by 
$q_{ab} = \frac{1}{N}\sum_{i=1}^N s^a_i s^b_i$ \cite{mezard2009}, where $s_i = 2x_i - 1$.
We show computed histograms of overlaps of local minima 
for the given random and industrial instances in Figs.~\ref{fig:uf_overlap} 
and~\ref{fig:semi_overlap}. 

The overlap distributions of random and structured problems exhibit distinct behaviour with the random instance having
the majority of states at zero overlap values. This property is not explicitly shown by the DG visualization.
It is implied, however, by the very large saddle clusters.
Compared to the random instance, the local minima of the factoring problem at $E \le 2$ are closer to each other
without showing evidence of a gap in the overlaps \cite{gamarnik2021}.
Thus, IMs can be challenged by different landscape features depending on the problem class,
suggesting a strong algorithmic need for specialisation.

\begin{figure*}[t]
    \includegraphics[width=0.86\linewidth]{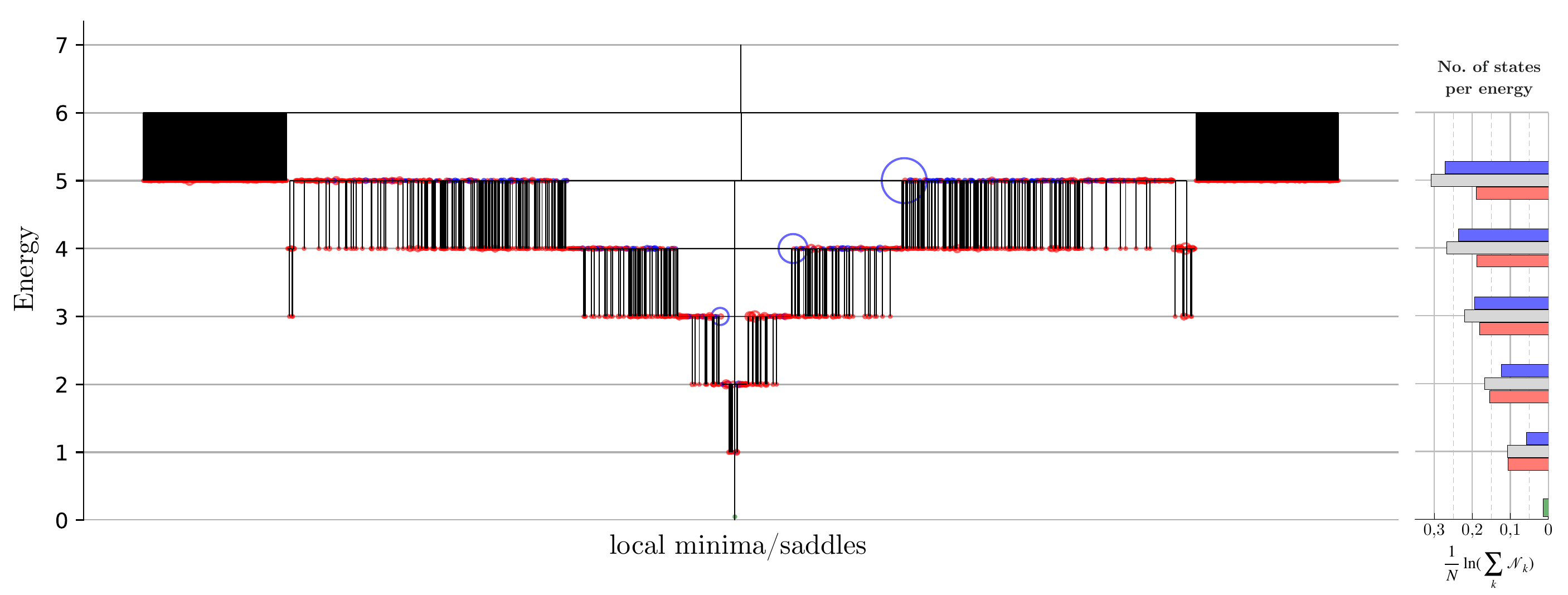}
    \caption{\label{fig:map2_uf920}
        QUBO mapping landscape (KZFD-BG) truncated to the stable states with 
        $E \leq 5$ of a 3-SAT instance in Fig.~\ref{fig:map0_uf920}.
        Problem size: 50 (native) + 136 (auxiliary) variables. QUBO factor $F=1$. 
        Total number of local minimum/saddle clusters is $1377$; 2 global minimum states.}
\end{figure*}
\begin{figure}[ht]
    \centering
     \begin{subfigure}[b]{1\linewidth}
         \centering
         \includegraphics[width=\linewidth]{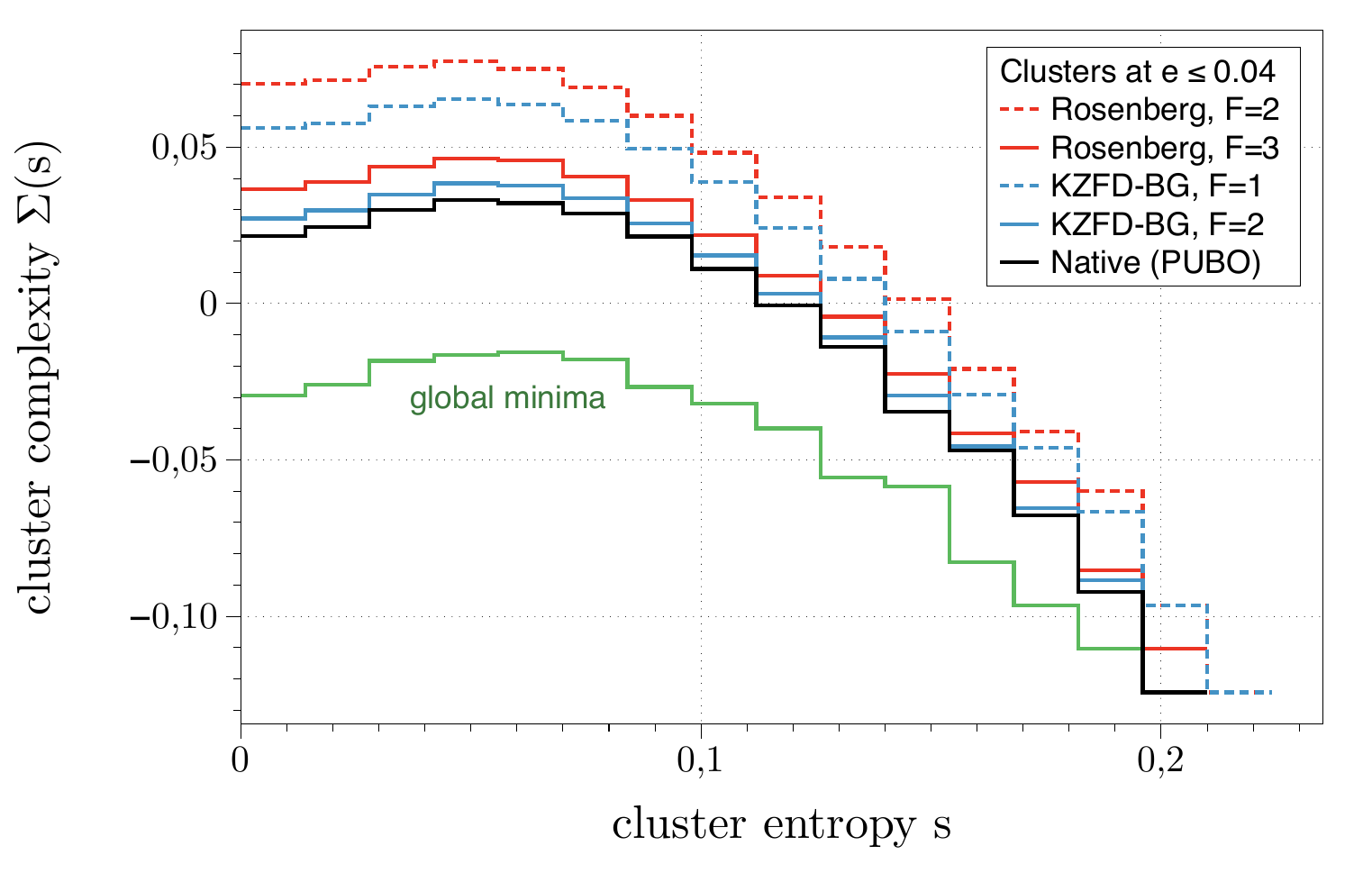}
         \caption{}
         \label{fig:mappings_comparison_1}
     \end{subfigure}
     \hspace{0.0\linewidth}
     \begin{subfigure}[b]{0.75\linewidth}
         \centering
         \includegraphics[width=\linewidth]{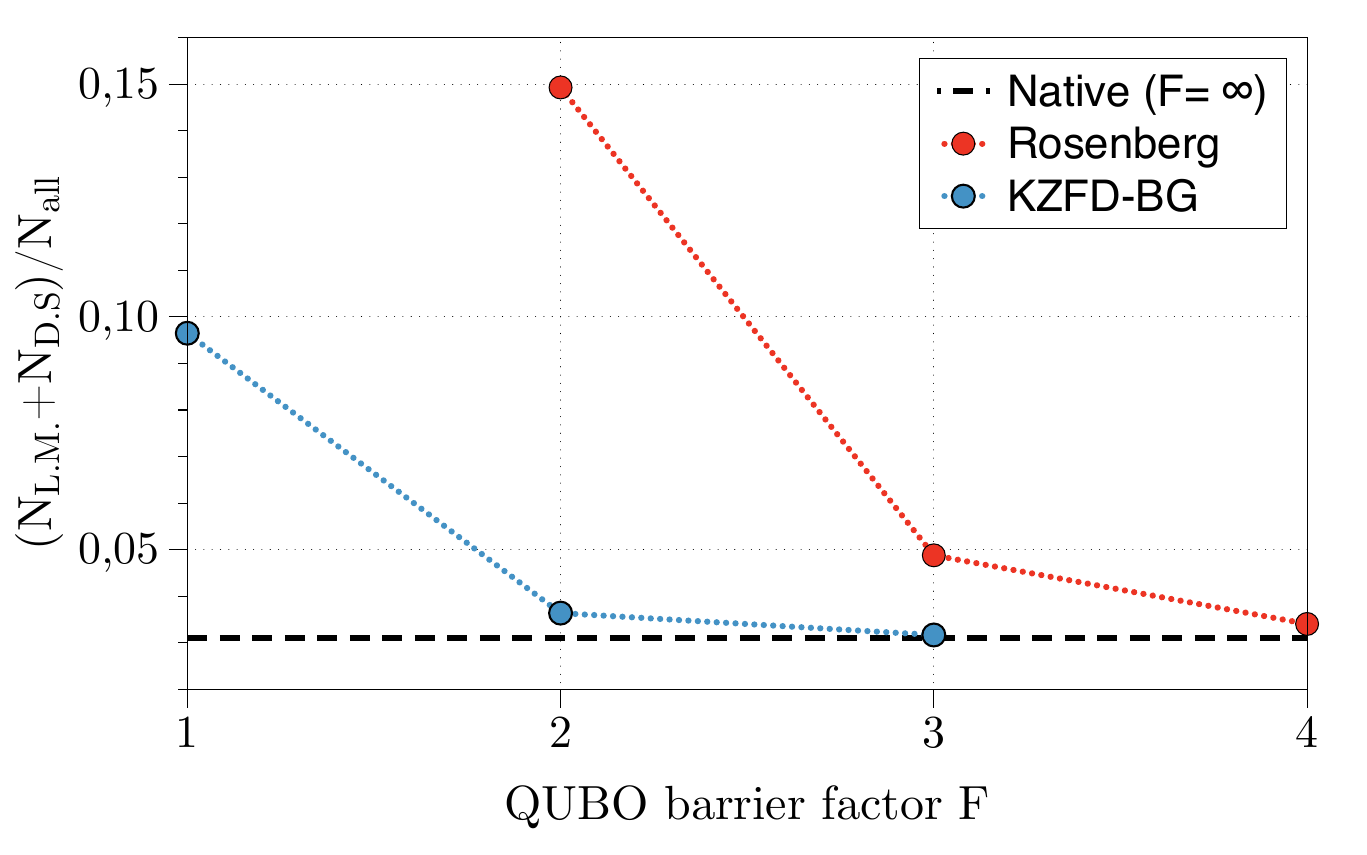}
         \caption{}
         \label{fig:mappings_comparison_3}
     \end{subfigure}
    \caption{\label{fig:mappings_comparison}
        (a) Local minimum and disconnected saddle cluster complexities
        vs cluster entropies for the native (PUBO) and QUBO landscapes 
        at energies $E\le 2$ for 500 sampled SATLIB 3-SAT problems of size $N=50$.
        (b) The ratio of the number of local minima and saddles disconnected from the 
        global minimum to all sampled stable states.
    }
\end{figure}

\subsection{QUBO mappings of 3-SAT \label{sec:qubo_3sat}}

In this section we study energy landscapes of QUBO mappings of 3-SAT using the DGs 
construction extension introduced in Sec.~\ref{sec:dg_qubo_extension}. We explicitly state the $F$ factor when 
mappings are compared with each other. We say that $F = \infty$ when all auxiliary variables are probed,  
essentially recovering the native (PUBO) landscape. At a given $F$ the ``effective'' barrier definition is illustrated
in Fig.~\ref{fig:qubo_factor}.

In Fig.~\ref{fig:map2_uf920} we plot a KZFD-BG QUBO mapping landscape of the instance from 
Fig.~\ref{fig:map0_uf920} truncated to the subspace of energies $E\le 5$. 
The QUBO barrier factor was chosen to be $F = 1$, 
meaning that at every step $\mathbf{x}_a \to \mathbf{x}_b$ 
only one QUBO auxiliary variable with $\Delta E > 0$ is allowed to be flipped 
in order to overcome the QUBO barriers between the native states $\mathbf{x}$.
One can observe the following features of the QUBO landscape of 3-SAT:
\begin{itemize}
    \item The connectivity of states is drastically reduced with large saddle point clusters of the PUBO landscape
    shattered into multiple disconnected saddles or local minima in the QUBO landscape (in total $1377$ clusters). 
    This has a direct negative effect on the ability to find 
    global minima for the local search heuristics at low temperatures/noise.
    \item The global minimum cluster (which consists of 2 states for this instance) is preserved, but only 
    $18$ compared to the original $206$ configurations were found to be connected from energy $E = 1$ towards the solution. 
    In other words, blue clusters have become red clusters.
    The same behaviour is observed higher in energies, i.e. local search faces new energy 
    barriers \textit{in addition} to entropic barriers. 
\end{itemize}

In order to highlight the necessity to carefully approach mapping into Ising hardware, 
we would like to directly compare the QUBO mappings introduced in Sec.~\ref{sec:locality_reduction}
and described in detail in App.~\ref{appx:quadratization} to each other
from the perspective of connectivity of states (clustering) at different values of the QUBO factor $F$.
We use 500 SATLIB instances of size 50 with 218 clauses 
(\textit{uf50 501-1000}) to accumulate statistics from sampled DGs. 
On average, our QUBO mapping scheme of variable substitution and 
penalty terms introduced $138\pm 4$ auxiliary variables.

In Fig.~\ref{fig:mappings_comparison_1} we show the histogram of the number of local minimum/disconnected saddle clusters 
$\mathcal{N}(s) \equiv \exp{(N\Sigma(s))}$ of size $S \equiv \exp{(Ns)}$ for KZFD-BG and Rosenberg mappings, 
and for the native space. The parameter $\Sigma(s)$ is usually referred to as cluster \textit{complexity}, while 
$s$ is the cluster \textit{entropy} \cite{mezard2005a}.
We consider the states sampled at energies $e = E/N \le 0.04$. $\Sigma(s)$ is computed as the logarithm
of the number of clusters of entropy $s$ averaged over 500 used instances.
As mentioned in Sec.~\ref{sec:dg_bs_extension}, the distinct clusters are sampled with the GWL algorithm,
while the cluster entropy estimations are improved further by BFS.
At the given energy levels we never reached the limit of BFS ($10^7$ states), 
which means that the size of every discovered cluster was exactly refined with BFS.
As discussed in App.~\ref{appx:dg_convergence}, we also made sure that the GWL sampling histogram 
was uniform for every mapping, and that on average every local minimum had approximately 
the same number of visits.

We explicitly plot the global minima ($E=0$) distribution as a sanity check: 
the RSB theory predicts its maximum value in the thermodynamic limit being at
the entropy $s\approx 0.06$, while the curve itself should be below 0 complexity when the clause-to-variable ratio 
of 3-SAT is above the phase transition value $4.267$ (we have $4.36$) \cite{montanari2008}. 
Both features are present for our sampled data.

We observe shattering of the native landscape clusters by the Rosenberg mapping to be stronger 
than that of the KZFD-BG mapping. This result can be interpreted as follows. 
On average, in order to transition (overcome the barrier) 
from state $\mathbf{x}_a$ to state $\mathbf{x}_b$ having the same energy, 
the Rosenberg mapping needs to overcome barriers for at least $F_{\mathrm{Ros}}$ auxiliary variables, introduced by quadratization, 
while KZFD-BG would safely transition after passing only $F_{\mathrm{KZFD}} < F_{\mathrm{Ros}}$. As a result, $F$ can be seen 
as a measure of the ruggedness or shattering of the quadratized optimization landscape induced by the mapping.

As displayed by the histogram in Fig.~\ref{fig:mappings_comparison_1}, 
both QUBO mappings feature large clusters that are not accounted for 
in the PUBO case. These are the native space saddle clusters connected to the global minimum (thus not shown 
on the PUBO histogram), which were transformed by quadratization to either local minima or disconnected saddles.
In Sec.~\ref{sec:easy_and_hard_dg} we discussed the value of connected 
saddle points at low temperatures/noise  for finding global minima using IMs. 
QUBO mappings, thus, can transform saddle points into local minima effectively impeding the descend in energy.
The ratio of local minima/disconnected saddles to all stable states is shown in Fig.~\ref{fig:mappings_comparison_3}.
With increasing $F$ we approach the native landscape faster for the KZFD-BG mapping than for the Rosenberg mapping. 
This suggests a potential algorithm for IMs that are forced by hardware to use quadratization methods. With sufficient
exploration of the auxiliary space, it is possible to recover the native (PUBO) landscape geometry and thereby 
benefit from the reduced number of local minima/disconnected saddles, provided that the costs 
of specific hardware implementations do not outweigh such benefits in terms of time-to-solution/energy-to-solution 
metrics.

\begin{figure}[ht]
    \includegraphics[width=1\linewidth]{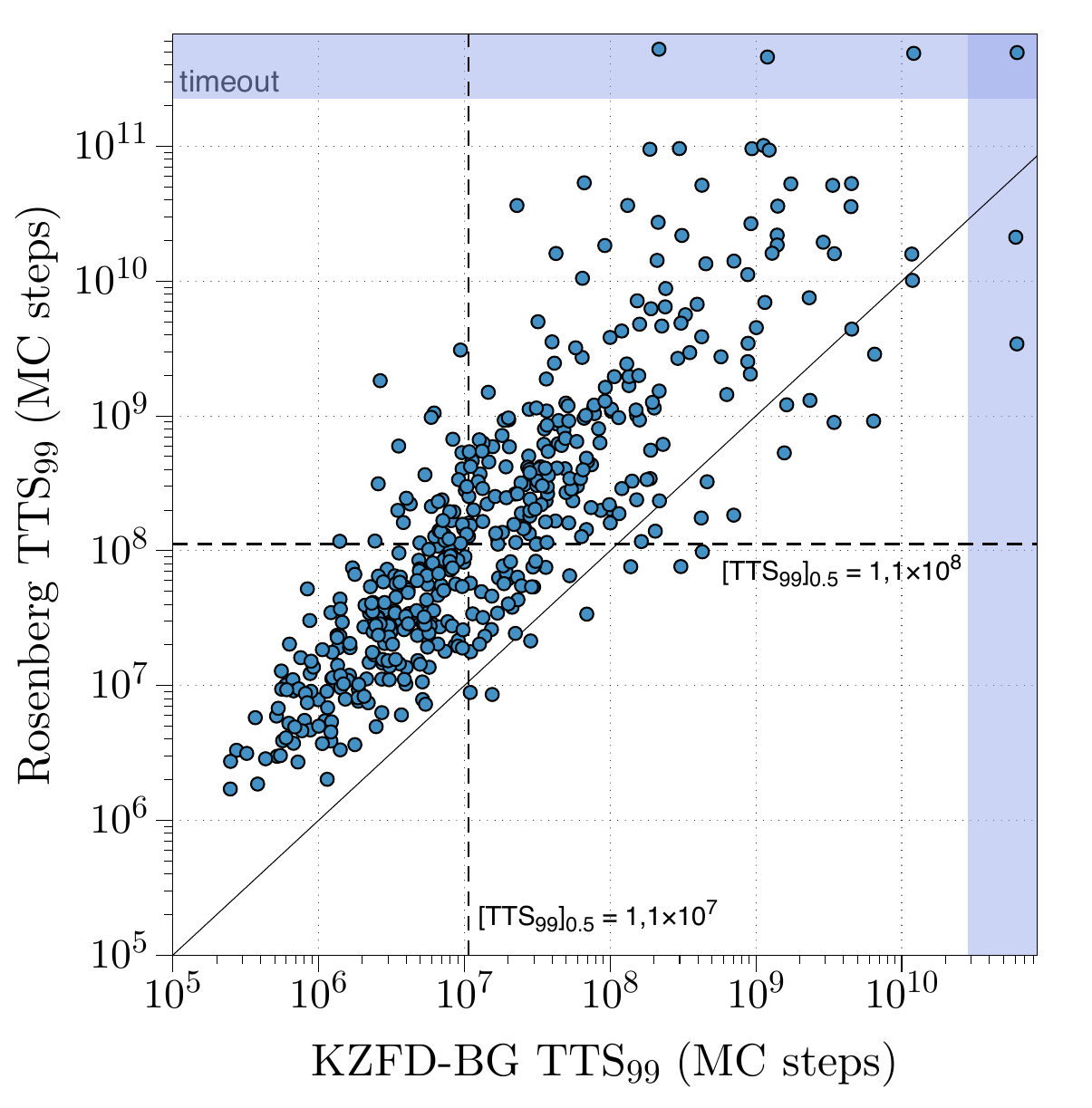}
    \caption{\label{fig:TTS_qubo_comparison}
        TTS$_{99}$ of simulated annealing for the QUBO mappings: non-term-wise  Rosenberg 
        and KZFD-BG. Instances SATLIB \textit{uf50 500-1000}. 
        The timeouts are different due to distinct optimal number of sweeps 
        (see App.~\ref{appx:sa_ho_opt}); 
        the solid line denotes the equality of TTS$_{99}$;
        the dashed lines indicate the medians of TTS$_{99}$.
    }
\end{figure}

With our analysis we would also like to highlight the effect of 
choosing quadratization methods on the performance of solvers. While all such methods preserve global minima, the 
geometry of the configuration space changes, thereby drastically decreasing the 
local search efficiency in terms of the time-to-solution pre-factor and empirical scaling with the problem size 
\cite{perdomo2019, hizzani2023, sharma2023, bybee2023}.
To support the observed energy landscape advantage of KZFD-BG mapping over Rosenberg,
we performed simulated annealing for a collection of SATLIB 3-SAT problem instances in Fig.~\ref{fig:TTS_qubo_comparison}.
The advantage of the KZFD-BG mapping clearly exhibits itself 
in the solver performance giving smaller TTS$_{99}$ for the majority of instances 
($\approx 96\%$). For this problem size of $N=50$ 
we observe an order of magnitude improvement of the median time-to-solution.
As a result, we would like to distinguish the energy landscape geometry features of different individual problem instances 
from the features of quadratization methods. In the former case, the details of the energy landscape 
result in the spread of computational hardness as we showed in the Fig.~\ref{fig:pubo_50_tts_hist}. 
In the latter case, the QUBO mappings can lead to orders of magnitude penalties on 
performance for any problem instance (comparing Figs.~\ref{fig:pubo_50_tts_hist} 
and ~\ref{fig:TTS_qubo_comparison}).
We refer to App.~\ref{appx:sa_details} for SA implementation details,
including hyperparameter optimization, error estimation, and timeout definition. 
In App.~\ref{appx:sa_ho_opt} we also test a larger problem size demonstrating increasing advantage of PUBO  
over QUBO, as well as of one QUBO mapping over the other, even though their corresponding QUBO
embedding size is the same. 

\section{Conclusion}
In this work we have suggested methods to sample disconnectivity graphs of degenerate combinatorial 
optimization problems, while also introducing extensions for quadratic embeddings motivated
by hardware constraints of Ising machines. DGs have proven to be able to visually capture energy landscape
properties of instances with different structure (industrial and random), hardness, and order 
(quadratic and higher order). To characterize clustering/ruggedness of the configuration
space arising from locality reduction, we have introduced a new method, QUBO factor $F$. 
From this perspective we have discussed the reasons behind 
observed experimental performance gap between different QUBO mappings, as well as between QUBO and PUBO. 

The directions for future work include investigating other definitions
of neighbourhoods beyond the simple bit-flip in order to visualize and gain intuition into how optimization 
landscapes are perceived by different local (or non-local) search routines. For example, isoenergetic cluster moves
\cite{houdayer2001} allow solvers to make large Hamming distance steps, 
defining a different neighbourhood for each configuration, 
thus a new DG with distinct connectivity of states. 
Moreover, understanding of the energy landscape geometry is of great importance in a variety of fields
ranging from inference and learning in energy-based models \cite{lecun2006} 
to attractor dynamics and storage capacity in associative memories \cite{hopfield1982, krotov2016}.
One application example is non-equilibrium inhomogeneous sampling methods \cite{mohseni2021, mohseni2023} 
which essentially modify energy barriers reducing the hardness of sampling of high-quality and diverse solutions.

Other embedding methods motivated by the available connectivity topology or the bit-precision requirements 
of the Ising hardware constitute a complementary problem  which can also be studied with the 
methods of this work. Finally, the distinct properties of auxiliary variables imply the 
possibility to introduce adaptive algorithms leveraging the specific
native-auxiliary interactions within the constrains of Ising machines.

\section{Acknowledgements}
This material is based upon work supported by the Defense Advanced Research Projects Agency (DARPA) 
under the Air Force Research Laboratory (AFRL) Agreement No.~FA8650-23-3-7313. 
The views, opinions, and/or findings expressed are those of the authors and should 
not be interpreted as representing the official views or policies of 
the Department of Defense or the U.S. Government.
The authors gratefully acknowledge computing time on 
the supercomputer JURECA \cite{JURECA} Forschungszentrum Jülich under grant no.~``optimization''.
We also gratefully acknowledge the generous funding of this work under NEUROTEC II 
(Verbundkoordinator / Förderkennzeichen: Forschungszentrum Jülich / 16ME0398K) 
by the Bundesministerium für Bildung und Forschung. 

\appendix

\section{Sparsification by auxiliary variables \label{app:sparsification}}
Let's assume that due to hardware limitations we are unable to support 
full interaction connectivity of a variable $x_1$ of the PUBO function in Eq.~\ref{eq:PUBO}.
Due to its multilinear form, we can split the interactions of $x_1$, i.e.
\begin{equation}
    f(\mathbf{x}) = A(x_{i_a}, \dots, x_{k_a})x_1 + B(x_{i_b}, \dots, x_{k_b})x_1\,,
    \label{eq:PUBO_split}
\end{equation}
where $A$ and $B$ are independent functions. $x_1$ in the second term 
can be substituted by an auxiliary variable $y$ with the introduction of a penalty as follows:
\begin{equation}
    g(\mathbf{x}, y) = Ax_1 + By + P(x_1 + y - 2x_1y)\,,
    \label{eq:PUBO_split_y}
\end{equation}
which obeys $f(\mathbf{x}) = \min_y g(\mathbf{x}, y)$ as in locality reduction methods 
if $P \ge |B|$.
As a result, the local search move $(x_1, y) = (0, 0) \to (\bar{x}_1, \bar{y}) = (1, 1)$ can be 
made with single flips through a higher energy barrier $A+|B|$ than in the denser 
original formulation (see Tab.~\ref{tab:PUBO_split_y}).

\begin{table}[ht]
	\centering
	\caption{\label{tab:PUBO_split_y} Sparsification truth table}
	\begin{tabular}{cccc}
        \toprule
        $x$ & $y$ &  $g(\mathbf{x}, y)$ & $f(\mathbf{x})$\\
        \midrule
        0 & 0 & 0 & \multirow{2}{5em}{\centering 0} \\
        0 & 1 & $B + |B|$ &  \\
        1 & 0 & $A + |B|$ & \multirow{2}{5em}{\centering $A + B$} \\
        1 & 1 & $A + B$ &  \\
        \bottomrule
	\end{tabular}
\end{table}

\section{3-SAT to QUBO mappings \label{appx:quadratization}}
In this section we describe in detail the mappings of 3-SAT problems 
formulated as conjugate normal forms (CNF) to quadratic pseudo-boolean functions (QUBO).
\paragraph{Notation:} $x$ are boolean variables, $l$ are literals that stand for either $x$
or its negation $\bar{x} \equiv 1 - x$, $y$ are boolean auxiliary variables.

The problem of maximizing the number of satisfied clauses of size $k = 3$ 
is reformulated as a minimization problem of a third order pseudo-boolean polynomial 
of literals as follows (inverting the expression and using De-Morgan law):

\begin{eqnarray}
    &&(\bar{l}_{1_1} \vee \bar{l}_{2_1} \vee \bar{l}_{3_1})\wedge (\bar{l}_{1_2} \vee \bar{l}_{2_2} 
        \vee \bar{l}_{3_2})\wedge\dots \nonumber\\ 
    &&\rightarrow l_{1_1}l_{2_1}l_{3_1} + l_{1_2}l_{2_2}l_{3_2} + \dots\,,
    \label{eq:pubo_from_3cnf}
\end{eqnarray}
where each $l_{a_i} = x_{a_i}$ or  $l_{a_i} = \bar{x}_{a_i}$, $i \in [1, 3M]$, $a \in [1, N]$.
A straightforward mapping of this expression to QUBO (quadratization) would be to introduce the Rosenberg penalties
for every term in the sum:
\begin{equation}
    \begin{aligned}
        &l_{1_1}l_{2_1}l_{3_1} + l_{1_2}l_{2_2}l_{3_2} + \dots  = 
            \min_{y \in \mathbb{B}} g(l(x), y)\,, \\
        &g(l(x), y) = \sum_{i \in [1, M]} y_i l_{3_i} + (3y_i - 2y_il_{1_i} - 2y_il_{2_i} + l_{1_i}l_{2_i})\,.
    \end{aligned}
    \label{eq:map1_def} 
\end{equation}
The validity of such quadratizaton (Eq.~\ref{eq:quadratization}) directly follows from the fact that auxiliary variables
$y_i$ are introduced independently for each term of Eq.~(\eqref{eq:pubo_from_3cnf}) 
and $l_1l_2l_3 = \min_y \left[yl_3 + (3y - 2yl_1 - 2yl_2 + l_1l_2)\right]$.

\subsection{Non-term-wise Rosenberg \label{appx:rosenberg_quadratization}}
In order to get the ``classic'' Rosenberg \cite{rosenberg1975} 
quadratizaton, we write the PUBO of Eq.~(\eqref{eq:pubo_from_3cnf}) for variables~$x$:
\begin{eqnarray}
    l_{1_1}l_{2_1}l_{3_1} &+& l_{1_2}l_{2_2}l_{3_2} + \dots = \sum_{i < j < k} S_{ijk}x_ix_jx_k \nonumber\\
        &+& \sum_{i < j} W_{ij}x_ix_j + \sum_i B_ix_i + C\,.
    \label{eq:pubo_x_from_l}
\end{eqnarray}
The pairs $x_m x_n$ in a set 
covering all terms of order 3 are substituted by auxiliary variables $y_{(mn)}$ with 
the addition of a penalties as in Eq.~(\eqref{eq:map1_def}):
\begin{eqnarray}
    &&\sum_{i < j < k} S_{ijk}x_ix_jx_k + \dots = \min_y  \sum_{(mn),k} S_{(mn)k}y_{(mn)}x_k + \nonumber\\
    &&+ \sum_{(mn)}P^R_{(mn)}(3y_{(mn)} - 2y_{(mn)}x_n - 2y_{(mn)}x_m + x_mx_n) \nonumber\\ 
    &&+ \dots [\le\mathrm{2nd\;order\;terms}]\,,
    \label{eq:map_ros_def}
\end{eqnarray}
where the lower bound penalty coefficients are now index-dependent \cite{babbush2013}:
\begin{eqnarray}
    &&P^R_{(mn)} \ge \max\left[\sum_k S^+_{(mn)k}, -\sum_k S^-_{(mn)k}\right] \nonumber\\
    &&S^+_{(mn)k} > 0,\; S^-_{(mn)k} < 0\,.
\end{eqnarray}

\subsection{Non-term-wise KZFD-BG \label{appx:kzfd_quadratization}}
Here we modify the Rosenberg mapping of Sec.~\ref{appx:rosenberg_quadratization} 
applying quadratization ideas of \cite{kolmogorov2004, freedman2005, boros2014}, where 
the positive and negative monomials get different penalty terms of Eq.~\ref{eq:KZFDBG} 
(here third order):
\begin{eqnarray}
    &&-x_1x_2x_3 = \min_y\left[ 2y - \sum_{i=1}^3 yx_i \right]\,, \nonumber\\
    &&x_1x_2x_3 = x_2x_3 -\bar{x}_1x_2 x_3 \nonumber\\ 
    &&= x_2x_3 + \min_y\left[2y - y\bar{x}_1- \sum_{i=2}^3 yx_i \right]\,.
\end{eqnarray}

Rearranging the summands in these equations, we get for the positive monomial:
\begin{equation}
    x_mx_nx_k \to yx_k + (y - yx_m - yx_n + x_mx_n)\,,
\end{equation}
and for the negative monomial:
\begin{equation}
   -x_mx_nx_{k'} \to -yx_{k'} + y - x_mx_n + (y - yx_m - yx_n + x_mx_n)\,.
\end{equation}
As a result, an arbitrary 3rd order pseudo-boolean function is quadratizatized as:
\begin{eqnarray}
    &&\sum_{i < j < k} S_{ijk}x_ix_jx_k + \dots = \min_y  \sum_{(mn),k} S^+_{(mn)k}y_{(mn)}x_{k} \nonumber\\
    &&+ \sum_{(mn),k'} S^-_{(mn)k'}(y_{(mn)}x_{k'} - y_{(mn)} + x_mx_n) \nonumber\\
    &&+ \sum_{(mn)}P^K_{(mn)}(y_{(mn)} - y_{(mn)}x_n - y_{(mn)}x_m + x_mx_n)\nonumber\\ 
    &&+ \dots [\le\mathrm{2nd\;order\;terms}]\,,
    \label{eq:map_kzfdbg_def}
\end{eqnarray}
where $S^+_{(mn)k'}, S^-_{(mn)k'}$ denote coefficients of positive and negative monomials.
The penalty parameters $P^K_{(mn)}$ are chosen as:
\begin{eqnarray}
    &&P^K_{(mn)} \ge \sum_k S^+_{(mn)k} - \sum_kS^-_{(mn)k}\,, \nonumber \\
    &&S^+_{(mn)k} > 0,\;S^-_{(mn)k} < 0\,.
    \label{eq:kzfd_penalty}
\end{eqnarray}
Indeed, for every auxiliary variable index $(mn)$ we have
\begin{eqnarray}
    &&g(x,y_{(mn)}) = \left(N^+ + N^-\right)y_{(mn)} 
    -|N^-|(x_mx_n - y_{(mn)}) \nonumber\\
    &&+ P^K_{(mn)}(y_{(mn)} - y_{(mn)}x_n - y_{(mn)}x_m + x_mx_n)\,,
\end{eqnarray}
where we defined 
\begin{eqnarray}
    &&N^+ \equiv \sum_{k} S^+_{(mn)k}x_k,\;
    N^- \equiv \sum_{k'} S^-_{(mn)k'}x_k \nonumber\\
    &&|N^+| = \sum_k S^+_{(mn)k},\;-|N^-| = \sum_kS^-_{(mn)k}\,.
\end{eqnarray}
As a result, $f(x) = \min_y g(x,y)$ due to Eq.~\ref{eq:kzfd_penalty} is guaranteed, as shown in Tab.~\ref{tab:map5_table}.
Compared to the Rosenberg mapping, non-term-wise KZFD-BG has smaller dynamic range second order interactions,
since that $2P^{R} > P^K$. 

\begin{table}[h]
	\centering
	\caption{\label{tab:map5_table} KZFD-BG QUBO mapping truth table for every 
    substituted pair $x_nx_m$ and an auxiliary $y_{(mn)}$.}
	\begin{tabular}{ccccc}
		\toprule
		$y_{(mn)}$     & $x_m$  & $x_n$   & $g(x,y_{(mn)})$ & $f(x)$ \\
        \midrule
        0 & 0 & 0 & 0 & \multirow{2}{5em}{\centering 0}\\
        1 & 0 & 0 & $N^+ + N^- + |N^-| + P^K_{(mn)}$ & \\
        0 & 1 & 0 & 0 & \multirow{2}{5em}{\centering 0}\\
        1 & 1 & 0 & $N^+ + N^- + |N^-|$ & \\
        0 & 1 & 1 & $-|N^-| + P^K_{(mn)}$ & \multirow{2}{5em}{\centering $N^+ + N^-$}\\
        1 & 1 & 1 & $N^+ + N^-$ & \\
		\bottomrule
	\end{tabular}
\end{table}

The same native variable pairs $x_i x_j$ are chosen for substitution
for both mappings in this work for fair comparison. Their choice is a result of a greedy 
(i.e. efficient) optimization algorithm choosing the most frequent variable pairs
which achieves significant reduction (possibly not the optimal 
\footnote{Finding a minimal set of pairs of variables to be substituted 
is in general an NP-Hard problem.})  
of the QUBO configuration space compared to the term-wise methods of Eq.~\ref{eq:map1_def}.

\section{Benchmarking methods \label{appx:sa_details}}
\subsection{Simulated annealing}
Simulated annealing (SA) \cite{kirkpatrick1983} is one of the simplest yet often powerful
physics-inspired heuristic algorithms which performs a 
MCMC (Markov Chain Monte Carlo) sampling following a predefined decreasing temperature schedule. 
There are two common MCMC transition probability rules \cite{katzgraber2011}:
the heat-bath, 
$p(\mathbf{x} \to \mathbf{x}') = \left[1+\exp(\beta \Delta E(\mathbf{x} \to \mathbf{x}'))\right]^{-1}$,
and the Metropolis-Hastings (used in this work), 
$$p(\mathbf{x} \to \mathbf{x}') = \min{\big[1,\,\exp(-\beta \Delta E(\mathbf{x} \to \mathbf{x}'))\big]}\,,$$
where $\beta = 1/T$.

The SA implementation we used to generate data for Fig.~\ref{fig:pubo_50_tts_hist} and Fig.~\ref{fig:TTS_qubo_comparison} 
follows an exponential temperature schedule $T(k) = T_{\rm{init}}\exp{(-\tau k/(N_{\rm{sweeps}}-1))}$, where
$\tau = \log(T_{\rm{init}}/T_{\rm{final}}$) and $k\in[0, N_{\rm{sweeps}}-1]$. At each $k$, 
we carry out one ``sweep'' over a permutation of $N$ ($N^{\rm{QUBO}}$ for the QUBO mapping) variables of the problem
applying the $p(x_i \to \bar{x}_i)$ rule. This results in a total $N\times N_{\rm{sweeps}}$ 
($N^{\rm{QUBO}}\times N_{\rm{sweeps}}$ for QUBO maps) MC steps for one SA run. The implementation of the SA used in this work 
is publicly available with the DG sampler code in \cite{sat-qubo-dg-sampler}.

\subsection{Error estimation}
Simulated annealing, being a heuristic probabilistic solver without guarantees, outputs a problem solution with a certain probability of 
success (POS) $\theta$. POS is defined as the number of successful runs $s$ out of all independent SA number of repetitions $N_{\rm{reps}}$. 
In this work, every such SA run gets its own random seed which results in an independent starting state and a sampling 
``trajectory'' followed. POS $\theta$ can exhibit strong instance-to-instance variation within a single problem class due to 
the distribution of problem hardness. Moreover, $\theta$ also depends on the algorithm hyperparameters and the 
problem size $N$ \cite{troels2014}.

The total effort $R_{p}$ of finding the ground state (or some predefined approximate solution) by a heuristic solver 
is commonly defined as the number of times the algorithm needs to be independently 
repeated in order to find a solution with probability $p$ (\%):
\begin{equation}
    R_{p}(\theta) = \frac{\log{(1-p)}}{\log{(1-\theta)}}\,.
\end{equation}
$R_{p}$ is then multiplied by a single SA run length to get the time-to-solution  metric 
TTS$_{p} = N\times N_{\rm{sweeps}}\times R_{p}$ (in MC steps). 
As a result, the wall-clock time can be readily estimated using one SA step cycle physical time of the CPU/GPU 
or an Ising-machine/dedicated hardware implementation. Due to the focus on the energy landscape 
geometry and the corresponding algorithmic penalties of the QUBO mappings, 
in this work we report all results in MC steps. 
Additionally, in Fig.~\ref{fig:TTS_qubo_comparison} we define an artificial 
``timeout'' value equal to TTS$_{99}^{\rm{timeout}} = N\times N_{\rm{sweeps}}\times R_{99}(0.5/N_{\rm{reps}})$.
This threshold value indicates instances with $\rm{TTS}_{p} > \rm{TTS}_{p}^{\rm{timeout}}$ having zero observed successful trials $s$.

We followed the works \cite{troels2014, hen2015} for the error estimation of the SA benchmarking data. 
Using the recorded number of successful trials $s$ from a number $N_{\rm{reps}}$ of independent SA repetitions, the 
probability distribution of the POS $\theta$ is modelled using the beta distribution: 
\begin{equation}
    \beta\left[1/2 + s, 1/2 + (N_{\rm{reps}} - s)\right]\,. 
    \label{eq:beta_pos}
\end{equation}
In order to generate the error-bars for a given value of interest $\mathcal{F}$ and a given set of instances $\mathcal{S}$, 
we use a simple bootstrapping method. A new set of instances $\mathcal{S}_i$ of the same cardinality as $\mathcal{S}$
is resampled with replacement from $\mathcal{S}$ $10000$ times. 
For each such instance $j$ in $\mathcal{S}_i$ the POS $\theta$ is sampled from the beta distribution of Eq.~\ref{eq:beta_pos}.
Finally, the statistics of $\mathcal{F}$ is obtained using the set $\mathcal{F}_i = \mathcal{F}(\{\theta_j\}_i)$.
For example, in Figs.~\ref{fig:sa_pubo_hpo}, \ref{fig:sa_qubo_hpo_50}, \ref{fig:sa_qubo_hpo_75} below
we report the mean and the standard deviation of the median TTS$_{99}$ using this bootstrapping method.
The same rule applies when we report the median of the ratios of TTS$_{99}$.

\subsection{Annealing hyperparameter optimization \label{appx:sa_ho_opt}}

\begin{figure}[ht]
    \centering
    \includegraphics[width=0.9\linewidth]{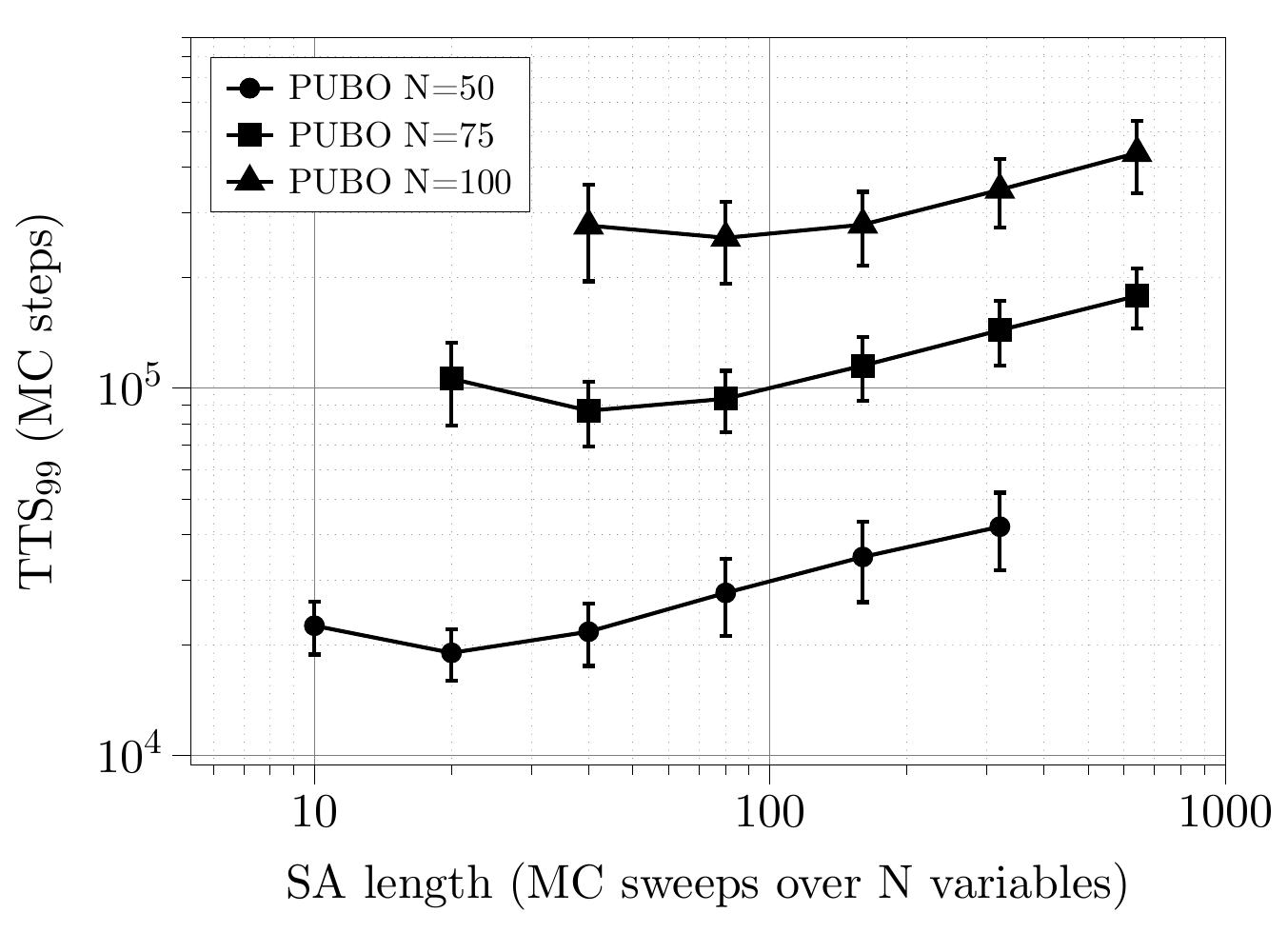}
    \caption{Hyperparameter optimization of the number of simulated annealing sweeps for PUBO. 
        $N_{\rm{reps}} = 5120,\,7680,\,10240$ of 50 instances each for $N=50,\,75,\,100$ respectively. 
        Mean and the standard deviation of the median TTS$_{99}$ estimated with bootstrapping.}
    \label{fig:sa_pubo_hpo}
\end{figure}

\begin{figure}[ht]
    \centering
    \includegraphics[width=0.9\linewidth]{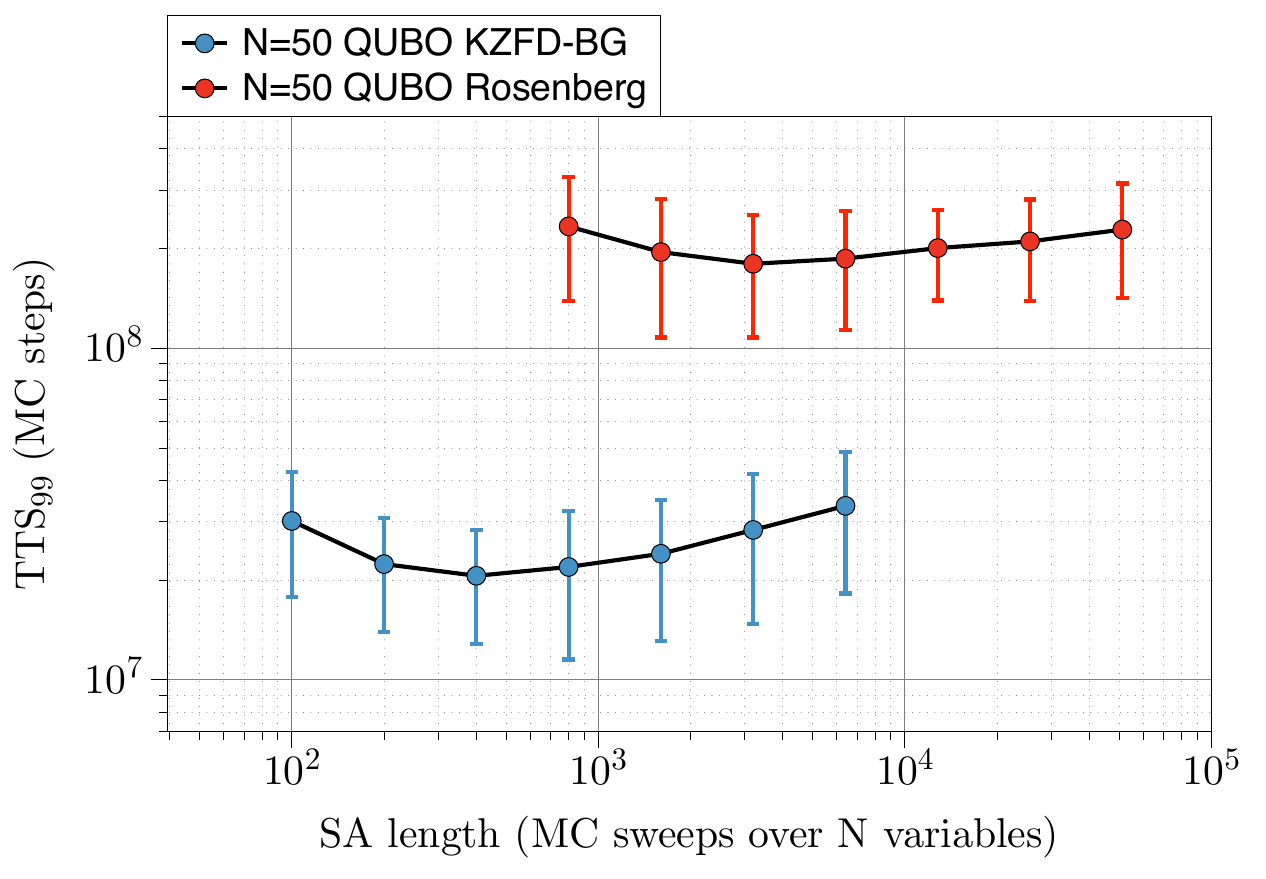}
    \caption{
        Hyperparameter optimization of the number of simulated annealing sweeps for the QUBO mappings. 
        Using 50 instances of size $N=50$, with mappings giving $N^{\rm{QUBO}} = 188 \pm 5$.
        $N_{\rm{reps}} = 20480$ for each instance. 
        Mean and the standard deviation of the median TTS$_{99}$ estimated with bootstrapping.}
    \label{fig:sa_qubo_hpo_50}
\end{figure}
\begin{figure}[ht]
    \centering
    \includegraphics[width=0.9\linewidth]{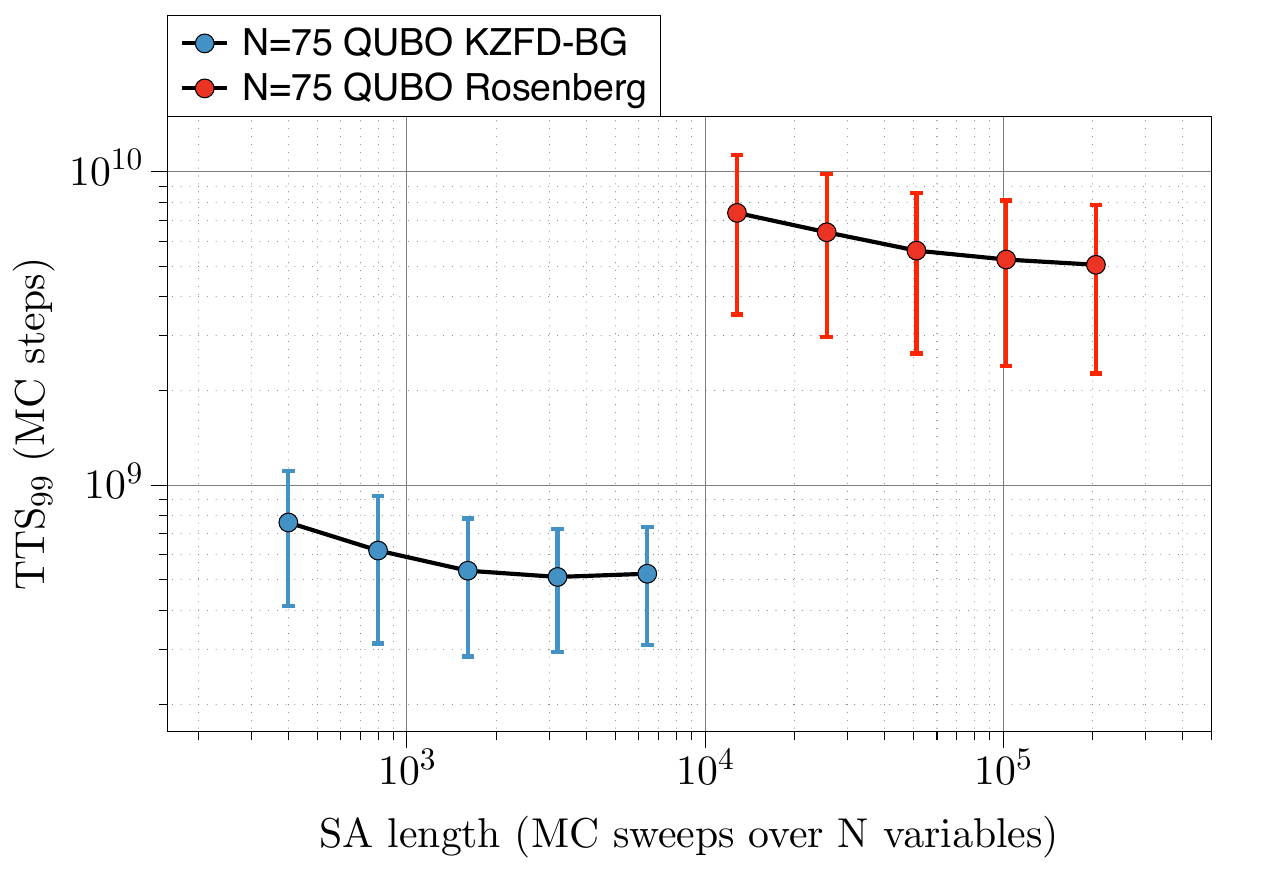}
    \caption{
        Hyperparameter optimization of the number of simulated annealing sweeps for the QUBO mappings. 
        Using 100 instances of size $N=75$, with mappings giving $N^{\rm{QUBO}} = 303 \pm 5$.
        $N_{\rm{reps}} = 81920$ for each instance. 
        Mean and the standard deviation of the median TTS$_{99}$ estimated with bootstrapping.}
    \label{fig:sa_qubo_hpo_75}
\end{figure}

The initial and final temperatures of SA set for the benchmarking of PUBO and QUBO mappings 
were chosen as: $T_i = 1.5$, $T_f = 0.1$. As a reminder, the minimum non-zero $|\Delta E|$ for the problems tested in this work
equals to $1$.
The chosen values of $T$ resulted on average in
the initial $0.615 \pm 0.005$, $0.634 \pm 0.010$, $0.523 \pm 0.012$  and 
the final $0.10 \pm 0.03$, $0.20 \pm 0.03$, $0.14 \pm 0.03$  
MCMC sampling acceptance rates for PUBO, QUBO KZFD-BG, and QUBO Rosenberg respectively.

PUBO and each QUBO mapping with the chosen temperature schedule
favour different $N_{\rm{sweeps}}$ for optimal performance. 
In order to facilitate fairer benchmarking, 
in Figs.~\ref{fig:sa_pubo_hpo}, \ref{fig:sa_qubo_hpo_50}, \ref{fig:sa_qubo_hpo_75}
we optimized this hyperparameter of SA. 
First, the PUBO performance at problem sizes $N=50$, $75$, $100$ is optimized in Fig.~\ref{fig:sa_pubo_hpo}
using first 50 satisfiable SATLIB 3-SAT instances at each size (\textit{uf(N) 1-50}).
Second, the QUBO performance optimization at problem size $N=50$ for the same 50 instances and at $N=75$ 
for all available 100 instances is shown in 
Fig.~\ref{fig:sa_qubo_hpo_50} and  Fig.~\ref{fig:sa_qubo_hpo_75}.

The mean and the standard deviation of the median TTS$_{99}$
were estimated using success probabilities obtained from $N_{\rm{reps}}$ experiment repetitions for each instance and each value
of $N_{\rm{sweeps}}$. 
As a result, we found the optimum $N_{\rm{sweeps}}$ and used these established values to generate results for 500 
instances \textit{uf50 501-1000} in Fig.~\ref{fig:pubo_50_tts_hist} and Fig.~\ref{fig:TTS_qubo_comparison} 
with increased number of the repetitions for even better statistics: $10240$ and $40960$ respectively.

Finally, we note the scaling differences between PUBO and QUBO with the problem size. 
The change of median TTS$_{99}$ from $N=50$ to $N=75$ in PUBO is: 
$18400 \pm 900$ for instances \textit{uf50 500-1000} in Fig.~\ref{fig:pubo_50_tts_hist} to
$83000 \pm 12000$ for instances \textit{uf75 1-100}, i.e. $\approx 5$ times increase.
In QUBO it equals to 
$(1.08  \pm 0.14) \times 10^7$ and $(1.07  \pm 0.17) \times 10^8$ at $N=50$ (Fig.~\ref{fig:TTS_qubo_comparison}) and
$(5.1  \pm 2.2) \times 10^8$ and $(5.0  \pm 2.8) \times 10^9$ at $N=75$
for KZFD-BG and Rosenberg mappings respectively (Fig.~\ref{fig:sa_qubo_hpo_75}). 
As a result, the increase of the median TTS$_{99}$ with increasing problem size for both 
mappings is $\approx 47$ times.

We also estimated the TTS$_{99}$ ratio of the QUBO mappings, 
namely the median of $\rm{TTS}^{Ros}_{99}/\rm{TTS}^{KZFD}_{99}$.
The resulting medians of ratios are $9.3 \pm 0.7$ at $N=50$ and $15.8 \pm 2.4$ at $N=75$.
As a result, the following conclusions can be made:
\begin{itemize}
    \item the advantage of PUBO vs QUBO grows with $N$, 
        i.e. the scaling of PUBO is exponentially better than both considered QUBO mappings;
    \item the scaling advantage with growing $N$ when comparing two QUBO mappings is also observed; 
        however, reliable functional fitting of scaling and extrapolation to larger problem sizes 
        requires extensive testing of the mappings at $N > 75$ and is left for future work.
\end{itemize}

\section{Disconnectivity graphs sampler \label{appx:dg_sampler}}
\subsection{Code availability \label{appx:code_availability}}
The original code developed for this work uses GWL sampling described in Sec.~\ref{sec:methods}
and has the following output: GWL histogram of visits to basins of attraction and energy levels, 
sampled clusters degeneracies, symmetric matrix of energy barriers between clusters, 
local minimum states.
The information about the connectivity of clusters and their type (local minimum/saddle) 
we then derive from the barrier matrix during post-processing and DG construction. As input the program 
takes the conjugate normal form of a 3-SAT problem.

The examples of DG sampling hyperparameters that can be tuned are: 
number of parallel threads of sampling, total GWL steps per thread,
limit on the cluster exploration and breadth-first search limits. 
In principle, it is possible to tune all hyperparameters to optimize the sampling for 
a particular problem class. The 3-SAT (PUBO/QUBO) GWL sampling code with the
hyperparameters used in this work as well as the 
simulated annealing implementation following App.~\ref{appx:sa_details}
is available in a public repository at ref.~\cite{sat-qubo-dg-sampler}.
Extended disconnectivity graphs construction from the aforementioned data
sampled with the GWL algorithm is based on the functions from 
\textbf{pele} library \cite{pele} and can be made available upon reasonable request. 

\subsection{GWL sampling uniformity \label{appx:sampling_uniformity}}
\begin{figure}[ht]
    \centering
    \includegraphics[width=1\linewidth]{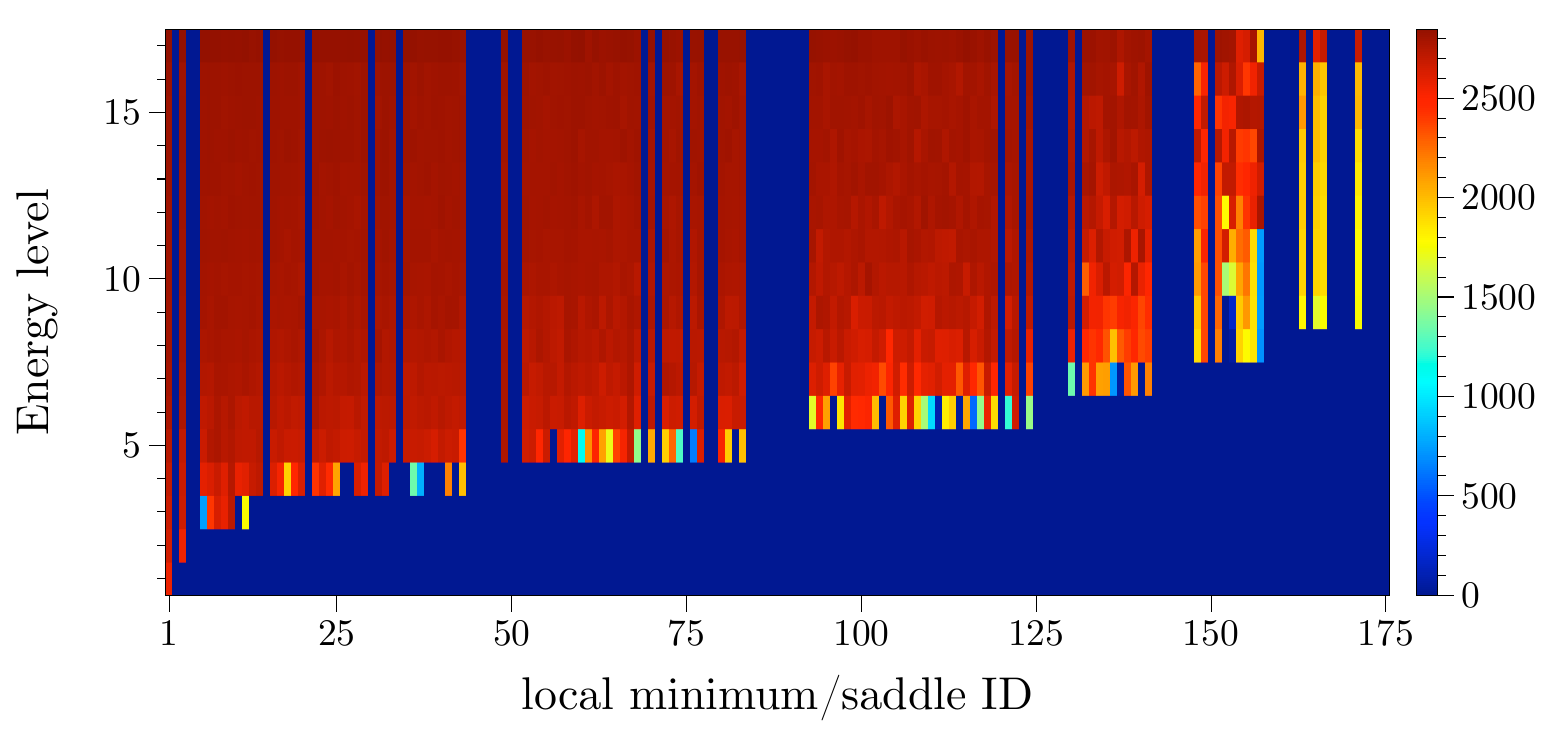}
    \caption{
        Sampling histogram of the 3-SAT instance in Fig.~\ref{fig:map0_uf920}.
        Total number of GWL steps is $4\times 10^6$.
        The visits to saddle clusters are counted towards their 
        corresponding local minima lower in energy, i.e. 
        saddles show zero visits in the histogram 
        (cf. Sec.~\ref{sec:dg_sampling}).
    }
    \label{fig:sampling_histogram}
\end{figure}

Generalized Wang-Landau (GWL) algorithm discussed in Sec.~\ref{sec:dg_sampling} aims to sample 
the configuration space as uniformly as possible. The uniformity of sampling is being tracked by the histogram 
$\theta_{l,k}$ \eqref{eq:gwl_histogram} of visits to a particular energy level $l$ and 
local minimum/saddle basin of attraction $k$. Fig.~\ref{fig:sampling_histogram} showcases one 
histogram example that we  observed while sampling the energy landscape for the DG construction
of Fig.~\ref{fig:map0_uf920}. We note that some of 
the independent clusters reported by the histogram can in fact be the same cluster with connections
between them not yet discovered during sampling. An extensive BFS search described in Sec.~\ref{sec:dg_bs_extension}
is employed in this work at post-processing to join such clusters together.

\begin{figure}[ht]
    \centering
    \includegraphics[width=1\linewidth]{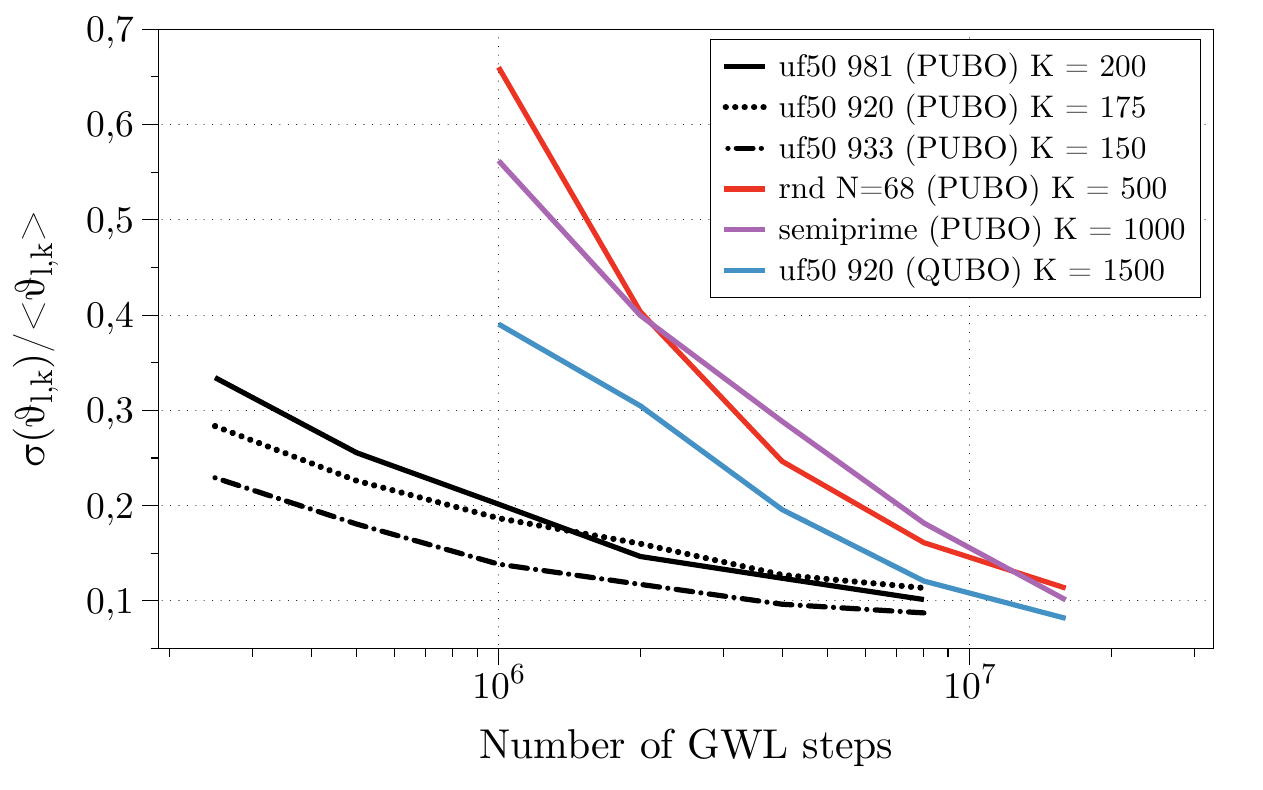}
    \caption{
        Relative standard deviation of the non-zero GWL histogram elements $\theta_{l, k}$ as 
        a function of the number of GWL sampling steps for every instance visualised with DGs
        (see histogram example in Fig.~\ref{fig:sampling_histogram}). 
    }
    \label{fig:std_histogram}
\end{figure}

By tracking the distribution properties of the histogram, one is able tune the hyperparameters 
of GWL sampling and/or observe its convergence. 
To verify our choices of hyperparameters, in Fig.~\ref{fig:std_histogram} we plot the relative standard
deviation (RSD) of $\theta_{l, k}$ as a function of the number of GWL MC steps. The maximum
histogram energy $E^L$ was chosen to be $16$ for \textit{uf50} instances in PUBO and QUBO, $18$ for the random
$N=68$ instance, and $17$ for the semiprime factoring instance.
Since the set of local minima/saddles during sampling is not fixed, in general 
the histogram may show temporary increases in its deviation due to the discovery of new clusters.
We have chosen RSD of $\approx 10$-$15\%$ for DG construction, which resulted in $4\times 10^6$ 
GWL steps for Fig.~\ref{fig:easy_and_hard_dg} and $1.6 \times 10^7$ for 
Figs.~\ref{fig:map0_n68uf1}, \ref{fig:map0_n68semi1}, and \ref{fig:map2_uf920}.

\subsection{Disconnectivity graphs convergence \label{appx:dg_convergence}}
\begin{figure}[ht]
    \centering
    \begin{subfigure}[b]{1\linewidth}
        \centering
        \includegraphics[width=\linewidth]{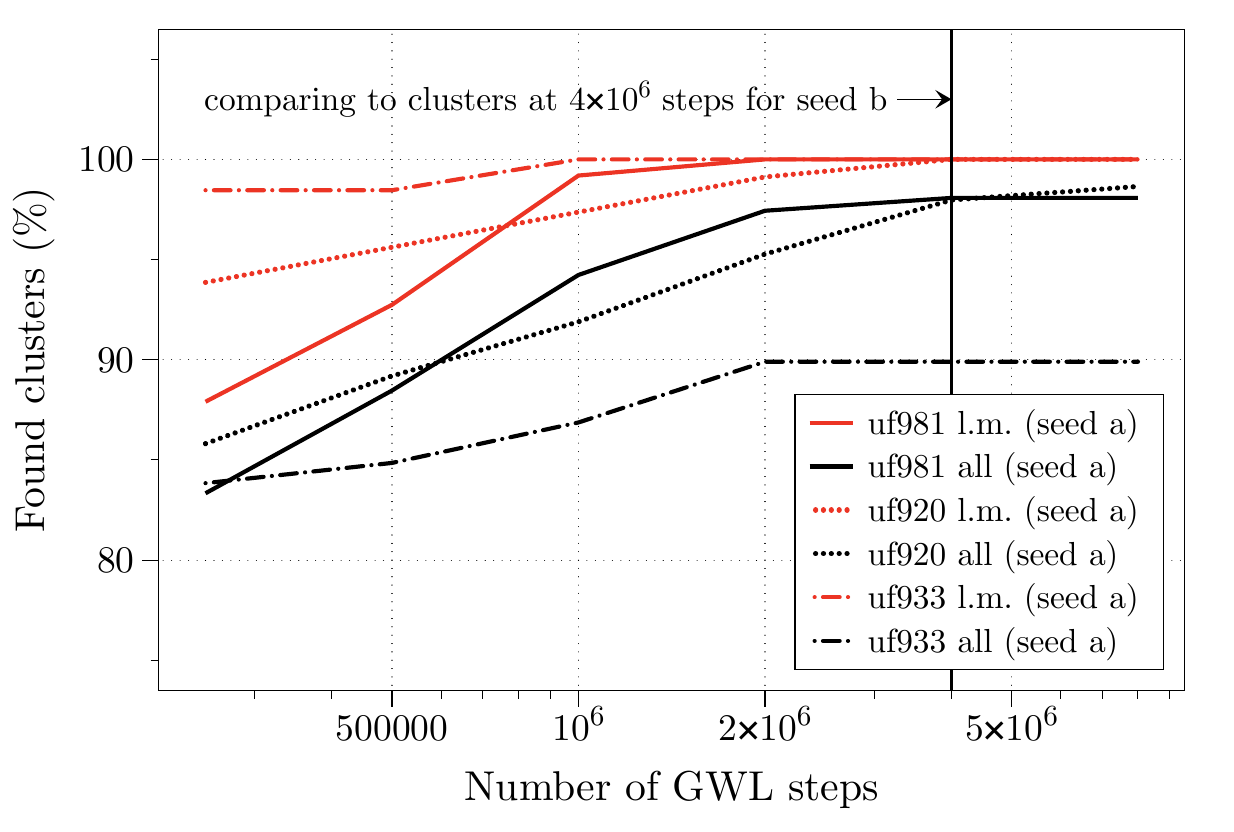}
        \caption{}
        \label{fig:clusters_saturation}
    \end{subfigure}
    \hspace{0.0\linewidth}
    \begin{subfigure}[b]{1\linewidth}
         \centering
         \includegraphics[width=\linewidth]{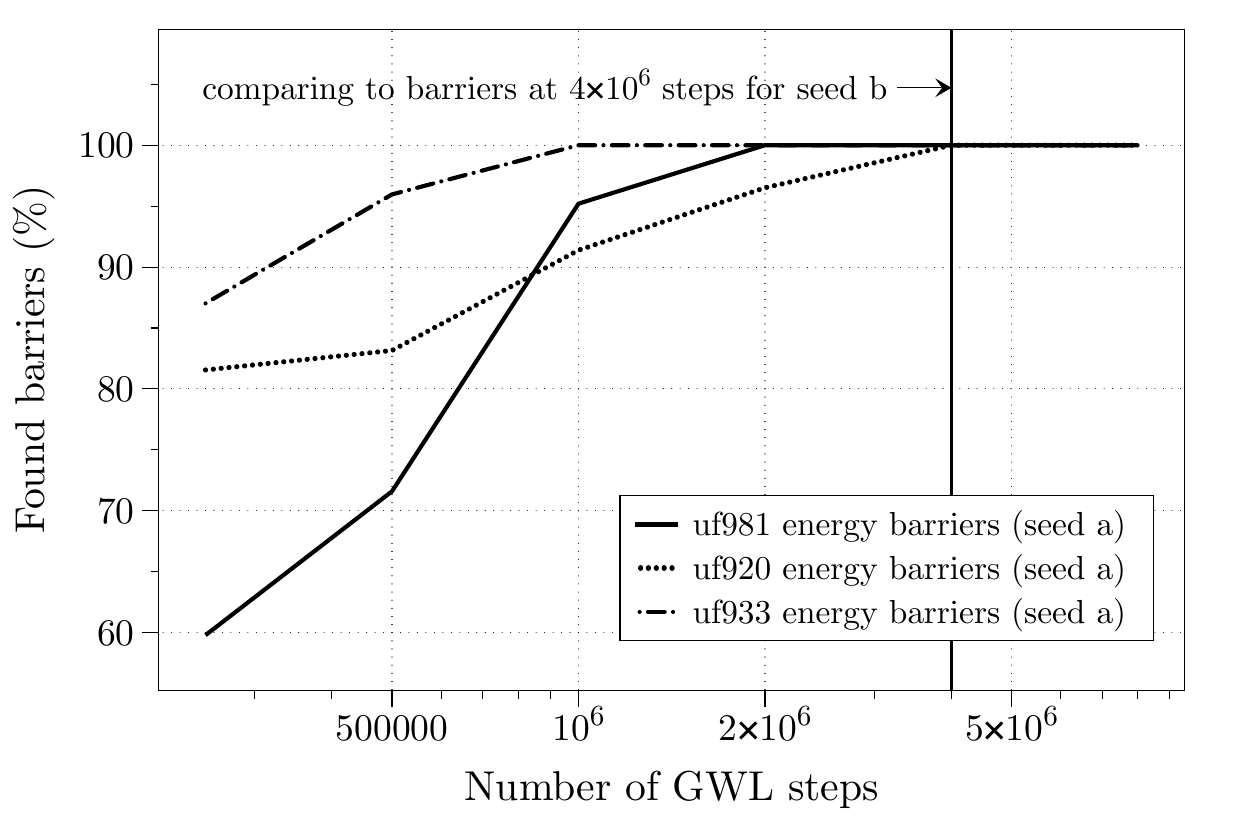}
         \caption{}
         \label{fig:barriers_saturation}
     \end{subfigure}
    \caption{\label{fig:saturation_1}
        The percentage of found number of local minimum clusters and all clusters 
        (local minima $+$ saddles) (a) and energy barriers (b) at different values of GWL sampling steps 
        and for an independent run (seed ``a'')
        with respect to the result used in this work ($4\times 10^6$ steps, seed ``b'') 
        in Fig.~\ref{fig:easy_and_hard_dg}.
    }
\end{figure}

Uniformity of the GWL histogram, while indicative of high quality of sampling,
is not a guarantee of DG construction accuracy. 
We have also tested the convergence (saturation) of the discovered local minimum/saddle clusters and 
of the energy barriers between them. 
In Fig.~\ref{fig:clusters_saturation}
we plot occurrence percentage of the cluster sets in an independent sampling run
at different numbers of GWL sampling steps to the sets in the sampling runs 
used for DGs in this work. These distinct runs differ only by a unique random seed choice.
In Fig.~\ref{fig:clusters_saturation} we observe that at $4$\,--\,$8\times 10^6$ all local minimum clusters are discovered 
with respect to the clusters obtained at $4\times 10^6$ in the GWL run used in this work, i.e. 
sampling saturated. 

The saddle clusters did not fully saturate due to the 
$10^7$ sampling limit, which restricted our ability to discover all stable states 
at very high energy levels and exactly match the clusters. 
However, this did not affect the accuracy of energy barrier construction, 
as shown in Fig.~\ref{fig:barriers_saturation}. 
Here we test how many of the $K'(K'-1)/2$ barriers between $K'$ local minima 
of the ``seed b'' runs have been reconstructed in the independent ``seed a'' runs. 
For every instance, we observe that the barriers between local minima 
saturate, indicating the reproducibility and accuracy of DGs construction.

In Fig.~\ref{fig:DG_uf_vs_semi}a, Fig.~\ref{fig:DG_uf_vs_semi}b, and Fig.~\ref{fig:map2_uf920} 
we found all discovered local minima at $E \le 4$, $E \le 3$, and $E \le 5$ respectively 
to coincide with an independent sampling run using $1.6\times 10^7$ GWL steps. At higher energy levels
our limits on the number of distinct clusters $K$ (given in Fig.~\ref{fig:std_histogram}) 
truncated different sets of local minima in independent runs; therefore, we do not 
compare the local minima found with seeds ``a'' and ``b''.
The barriers between the matched local minima were observed to $100\%$ match in 
Figs.~\ref{fig:DG_uf_vs_semi}a,b and $99.85\%$ match in Fig.~\ref{fig:map2_uf920}.

\subsection{Sampling complexity \label{appx:sampling_complexity}}
Sections \ref{sec:dg_sampling}-\ref{sec:dg_qubo_extension} describe a variety 
of primitives that were implemented in \cite{sat-qubo-dg-sampler} in order to construct DGs.
Above we reported the numbers of GWL MC steps that were used to obtain the data about local minima/saddles, 
as well as about the energy barriers between then. Each GWL step consists of the following routines, each
having its corresponding complexity.

When a new state is proposed, random descend is performed to establish the affinity to a particular basin of attraction.
Each step of the random energy descend requires the computation of $\Delta E$ of bit-flip neighbours.
If no negative $\Delta E$ is found, the worst case number of computations is $O(N)$,
assuming a sparse problem without scaling of the number of interactions for each variable.
When a plateau region is encountered during the descend, we perform a fixed predefined
number of exploration steps before terminating the descend. It is a hyperparameter and chosen to be $20$ 
in this work. We do not scale this number with the problem size, 
thus the complexity is also $O(N)$. Since the total number of descend steps scales as $O(N)$, 
the resulting complexity is $O(N^2)$.

When a saddle or a local minimum is identified, we need to either find an existing cluster it is connected to, 
or insert it as a new cluster to the set of all clusters. 
At each energy level we store a sorted set container of all so far discovered stable states.
Let us assume that there are exponentially many already found states, i.e. the container size is worst case 
$O(\exp{cN})$. The search and insertion into such a sorted container has complexity $O(\log(\rm{size}))$, i.e. $O(N)$.
Since we need to identify and insert $O(N)$ states, due to $O(N)$ possible energy levels 
and a fixed number of newly discovered states at each level, the overall complexity of search and insertion is $O(N^2)$.

Finally, the breadth-first search can be executed in our implementation 
when a new cluster is discovered, or at the end of GWL sampling to exactly calculate 
the the size of each cluster. In the former case, the limit is a hyperparameter, 
which in work was chosen to be 500. In the latter case, for each energy level 
we set the total limit on states to $10^7$ for the DGs of problems of native size $N=50$ 
(PUBO and QUBO) and $5\times 10^7$ for DGs of the problems of size $N=68$.
The number of required iterations depends on the degeneracy of the problem of interest.

In this work, our main focus was the accuracy of the introduced DGs sampling and 
construction method demonstrated for the chosen hyperparameter in the sections above.
Our machine (single thread of a CPU) took $\approx 20$\,--\,$30$ minutes for GWL sampling of DGs
in Figs.~\ref{fig:easy_and_hard_dg} ($4\times 10^6$ steps), 
$\approx 2$\,--\,$3$ hours in Figs.~\ref{fig:DG_uf_vs_semi}, 
and \ref{fig:map2_uf920} ($1.6\times 10^7$ steps), 
each having tracked tens of of millions of local minimum/saddle states. 
The program \cite{sat-qubo-dg-sampler} supports multitheading speedup due to 
parallel independent sampling of a single energy landscape. 
We leave the exhaustive hyperparameter optimization and 
benchmarking of our implementation of the method for future work.

\subsection{QUBO sampling details \label{appx:qubo_sampling}}
\begin{figure}[ht]
    \centering
    \includegraphics[width=0.97\linewidth]{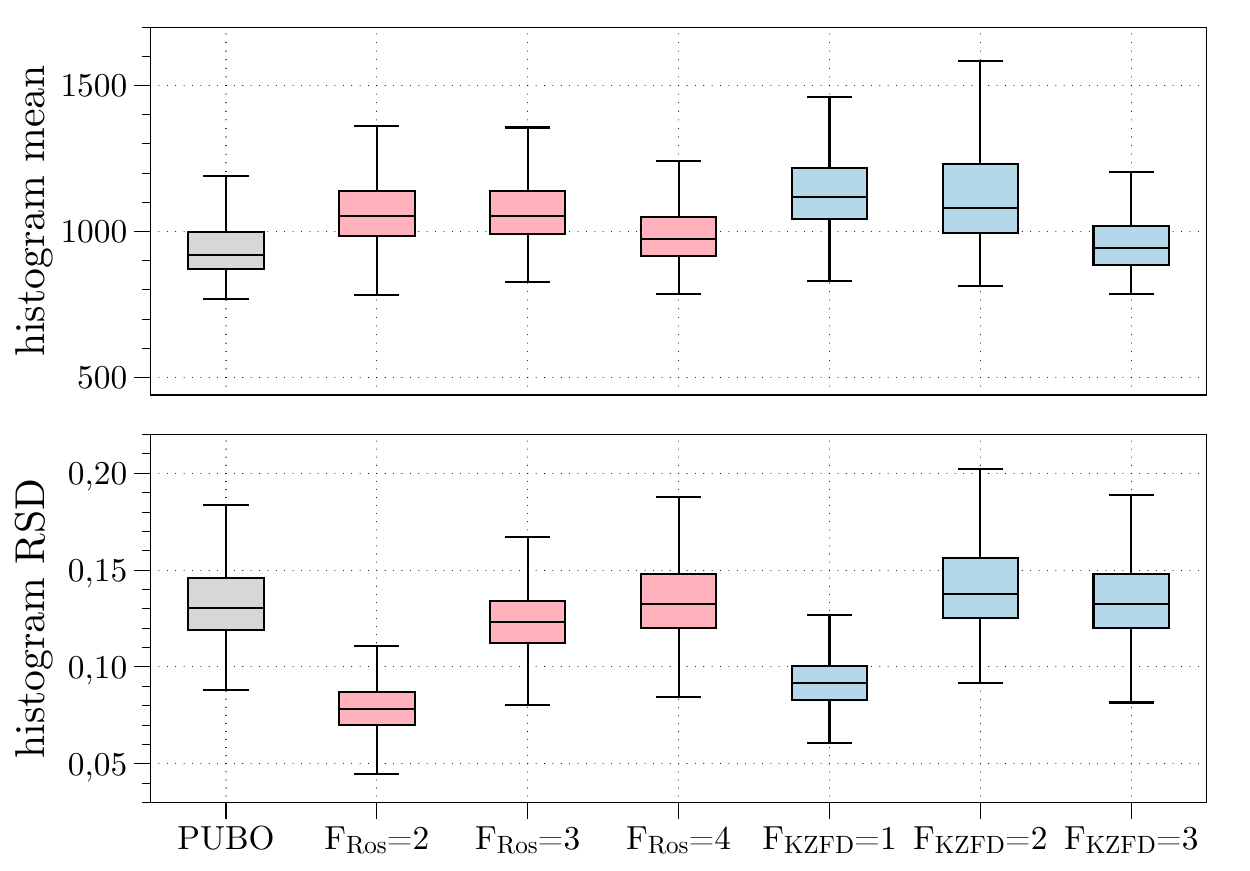}
    \caption{
        Statistics of the means and of relative standard deviations (RSD) of the non-zero GWL histogram elements 
        $\theta_{l, k}$ for instances and mappings in Fig.~\ref{fig:mappings_comparison}.}
    \label{fig:qubo_histograms_stats}
\end{figure}

For every QUBO mapping and every value of $F$ in Fig.~\ref{fig:mappings_comparison}, 
we have chosen the histogram size limits $K$ so that we are able to fit all
distinct clusters at the energy levels $E\le 2$. Next, the 
numbers of GWL sampling steps were chosen so that in each case we obtain good 
levels of histogram relative standard deviation ($\lessapprox 15\%$) and approximately 
equal number of absolute visits to each basin of attraction ($\approx 1000$). For most instances,
$K = 70$ was sufficient for PUBO, $K = 500,\,200,\,100$ for the Rosenberg mapping with $F = 2,\,3,\,4$ 
respectively,\,and $K = 400,\,150,\,80$ for the KZFD-BG mapping with $F = 1,\,2,\,3$ respectively.
We found the value of $N_{\rm{steps}} = K\times 10^4$ to satisfy the aforementioned requirements.
The resulting statistics of $\theta_{l, k}$ of instances in Fig.~\ref{fig:mappings_comparison} 
are shown in Fig.~\ref{fig:qubo_histograms_stats}.

\bibliography{main}  
\end{document}